\documentclass[iop,revtex4]{emulateapj}
\usepackage{verbatim}
\usepackage{pdflscape}
\usepackage{xcolor}

\definecolor{darkred}{rgb}{0.8,0.13,0.13}
\definecolor{darkblue}{rgb}{0.13,0.15,0.8}
\definecolor{medblue}{rgb}{0,0,0.5}
\usepackage{hyperref}
\hypersetup{
    pdfauthor={Xingxing Huang},
    pdftitle={CLASH: Extreme Emission Line Galaxies and Their Implication on Selection of High-Redshift Galaxies},
    pdfstartview=FitH,
    pdfstartview=Fit,
    pdfpagelayout=SinglePage,
    pdfpagemode=UseNone,
    colorlinks=true,
    linkcolor={darkred},
    citecolor={darkblue},
    urlcolor={medblue},
}

\begin{document}

\newcommand{\kms}{\>{\rm km}\,{\rm s}^{-1}}
\newcommand{\msol}{M_{\odot}}
\newcommand{\flux}{$\rm {ergs\ s^{-1}cm^{-2}}$ }
\newcommand {\oiii}{[\ion{O}{3}] $\lambda 5007$}
\newcommand {\oiiidouble}{[\ion{O}{3}] $\lambda \lambda 4959,5007$}
\newcommand {\oii}{[\ion{O}{2}] $\lambda 3727$}
\newcommand {\ha}{H$\alpha$ $\lambda 6563$}
\newcommand {\hb}{H$\beta$ $\lambda 4861$}
\newcommand {\lya}{Ly$\alpha$}
\newcommand {\lm}{$\lambda$}
\newcommand {\HST}{{\it HST}}
\newcommand {\etal}{et al.}
\newcommand {\ergscmA}{$\rm{{ergs\,s}^{-1}\,{cm}^{-2}\,{\AA}^{-1}}$}
\def\gtsim{\raisebox{-.5ex}{$\;\stackrel{>}{\sim}\;$}}
\def\arcsecpoint{\ifmmode ''\!. \else $''\!.$\fi}

\title{CLASH: Extreme Emission Line Galaxies and Their Implication on Selection of High-Redshift Galaxies}

\author{
Xingxing Huang\altaffilmark{1,2},
Wei Zheng\altaffilmark{2},
Junxian Wang\altaffilmark{1},
Holland Ford\altaffilmark{2},
Doron Lemze\altaffilmark{2},
John Moustakas\altaffilmark{3},
Xinwen Shu\altaffilmark{4,1},
Arjen Van der Wel\altaffilmark{5},
Adi Zitrin\altaffilmark{6,7},
Brenda L. Frye\altaffilmark{8},
Marc Postman\altaffilmark{9},
Matthias Bartelmann\altaffilmark{10},
Narciso Ben\'itez\altaffilmark{11},
Larry Bradley\altaffilmark{9},       
Tom Broadhurst\altaffilmark{12,13},    
Dan Coe\altaffilmark{9},
Megan Donahue\altaffilmark{14},
Leopoldo Infante\altaffilmark{15},
Daniel Kelson\altaffilmark{16},
Anton Koekemoer\altaffilmark{9}, 
Ofer Lahav\altaffilmark{17},
Elinor Medezinski\altaffilmark{2},
Leonidas Moustakas\altaffilmark{16},
Piero Rosati\altaffilmark{19},
Stella Seitz\altaffilmark{20},
Keiichi Umetsu\altaffilmark{21}
}

\altaffiltext{1}{CAS Key Laboratory for Research in Galaxies and Cosmology, Department of Astronomy, University of
Science and Technology of China, Hefei, Anhui 230026, China; e-mail:hxx@mail.ustc.edu.cn}
\altaffiltext{2}{Department of Physics and Astronomy, Johns Hopkins University, 3400 N. Charles Street, Baltimore, MD 21218}
\altaffiltext{3}{Department of Physics \& Astronomy, Siena College, 515 Loudon Road,Loudonville, NY, 12211, USA}
\altaffiltext{4}{CEA Saclay, DSM/Irfu/Service d'Astrophysique, Orme des Merisiers, 91191 Gif-sur-Yvette Cedex, France}
\altaffiltext{5}{Max-Planck Institute for Astronomy, K\"onigstuhl 17, D-69117, Heidelberg, Germany}
\altaffiltext{6}{Cahill Center for Astronomy and Astrophysics, California Institute of Technology, MS 249-17, Pasadena, CA 91125}
\altaffiltext{7}{Hubble Fellow}
\altaffiltext{8}{Steward Observatory/Department of Astronomy, University of Arizona, 933 N Cherry Ave, Tucson, AZ 85721-0065}
\altaffiltext{9}{Space Telescope Science Institute, 3700 San Martin Drive, Baltimore, MD 21208, U.S.A.}
\altaffiltext{10}{Leiden Observatory, Leiden University, P. O. Box 9513,2300 RA Leiden, The Netherlands}
\altaffiltext{11}{Instituto de Astrof\'isica de Andaluc\'ia(CSIC), C/Camino Bajo de Hu\'etor 24, Granada 18008, Spain}
\altaffiltext{12}{Department of Theoretical Physics, University of Basque Country UPV/EHU, Bilbao, Spain}
\altaffiltext{13}{IKERBASQUE, Basque Foundation for Science, Bilbao, Spain}
\altaffiltext{14}{Department of Physics and Astronomy, Michigan State University, East Lansing, MI 48824, USA}
\altaffiltext{15}{Departamento de Astrono\'ia y Astrof\'isica, Pontificia Universidad Cat\'olica de Chile, V. Mackenna 4860, Santiago 22, Chile}
\altaffiltext{16}{Observatories of the Carnegie Institution of Washington, Pasadena, CA 91 101, USA}
\altaffiltext{17}{Department of Physics and Astronomy. University College London, Gower Street, London WCIE 6 BT, UK}
\altaffiltext{18}{Center for Astrophysics and Space Sciences, University of California at San Diego. 9500 Gilman Dr., MC 0424, La Jolla, CA 92093, USA}
\altaffiltext{19}{Institute of Astronomy and Astrophysics, Academia Sinica, P. O. Box 23-141, Taipei 10617, Taiwan}
\altaffiltext{20}{Universit\"ats-Sternwarte, M\"unchen, Scheinerstr. 1, D-81679 M\"unchen, Germany}
\altaffiltext{21}{Institut f\"{u}r Theoretische Astrophysik, ZAH, Albert-Ueberle-Stra\ss e 2, 69120 Heidelberg, Germany}

\begin{abstract}                                                                           
We utilize the  CLASH (Cluster Lensing And Supernova survey with Hubble) observations of 25 clusters
to search for extreme emission-line galaxies (EELGs).
The selections are carried out in two central bands: F105W ($Y_{105}$) and F125W
($J_{125}$), as the flux 
of the central bands could be enhanced by the presence of [\ion{O}{3}]
$\lambda\lambda 4959,5007$ at redshift of $\sim 0.93-1.14$ and $1.57-1.79$, respectively.
The multi-band observations help to constrain the equivalent widths of emission lines.
Thanks to cluster lensing,
we are able to identify 52 candidates
down to an intrinsic limiting magnitude of 28.5 and to 
a rest-frame \oiiidouble\ equivalent width of $\simeq 3737$ \AA.
Our samples include a number of EELGs at lower luminosities that are missed in other surveys, 
and the extremely high equivalent width can be only found in such faint galaxies.
These EELGs can mimic the dropout feature similar to that of high redshift
galaxies and contaminate the color-color selection of high redshift galaxies when the S/N ratio
is limited or the band coverage is incomplete. 
We predict that the fraction of EELGs in the future high redshift galaxy selections cannot be neglected.
                                          
\end{abstract}

\section{Introduction}
The presence of extremely strong emission lines such as the \oiiidouble\ and \ha\ emission lines is one of the prominent spectral features in star-forming galaxies. 
Recently, a considerable number of star-forming galaxies with extraordinarily strong \oiiidouble\ \citep{straughn09, van11, atek11, smit14}
or \ha\ \citep{shim11,shim13} lines have been found. While some of these galaxies are identified spectroscopically
\citep{erb06,atek11,frye12}, the majority of them are found from broad-band photometry with a significant flux excess in one or more bands.
Utilizing \HST/Wide Field Camera 3 (WFC3) observations of the Cosmic Assembly Near-IR Deep Extragalactic Legacy Survey (CANDELS, \citealt{grogin11}),
\citealt{van11} (VDW11 hereafter) identified an abundant population of extreme emission line galaxies (EELGs) at redshift $z \sim 1.7$.
In some cases, the rest-frame equivalent widths (EWs) of such strong emission lines reach 2000 \AA\ or even higher.

Extremely strong emission lines can affect the spectral-energy-distribution (SED) fitting of broad-band
photometry \citep{schaerer09,atek11,shim11,labbe10,labbe12,stark13,fumagalli12}.
Their contributions may mimic the spectral feature of the Lyman break in high-redshift galaxies.
It is therefore possible that some high-redshift Lyman break galaxies (LBG) may actually be low-redshift EELGs
when the wavelength coverages or depths are limited.
Recently, the search for LBG has reached $z>9$,
and \HST\ plays the major role in this redshift range with the WFC3/IR instrument 
\citep{bouwens11,zheng12, coe13,ellis13,oesch13}.
UDFj-39546284 was first detected in the Hubble Ultra Deep Field (HDF09) with an $H_{160}$ band detection only \citep{bouwens11}.
The decrement between the F160W and F125W bands is larger than two magnitudes, thus suggestive of a $z \simeq 10$ candidate. 
Followup observations of the HUDF12 (GO 12498: PI Ellis) and 
CANDELS program \citep{grogin11} confirm that 
this substantial break in the SED is actually between the F160W and F140W bands \citep{ellis13,bouwens13a}, 
implying an even higher redshift. 
\citet{brammer13} analyzed deep WFC3 grism observations of UDFj-39546284
and found a 2.7$\sigma$ detection of an emission line at 1.599 $\micron$.
In the deep Keck observation, \citet{capak13} also found a 2.2$\sigma$ peak at the same wavelength. 
Both spectra suggest that UDFj-39546284 could be a strong \oiii\ emitter at $z\sim2.19$ or a strong \oii\ emitter at $z\sim3.29$.
Current deep near infrared observations are unlikely to determine the nature of this candidate. 
The presence of UDFj-39546284 suggests that
the possible contamination due to EELGs at lower redshift should be reexamined. 

In this paper, we report the search for EELGs at two redshift ranges in the CLASH cluster fields
to estimate the contamination to the selections of LBG. 
The CLASH program \citep{postman12} is a 16-band survey of 25 clusters between 0.2 and 1.6 \micron. 
The nominal limiting magnitude in the F160W band is approximately 27.2 ($5 \sigma$ detection limit). 
With the power of cluster lensing, some of target can reach an
 intrinsic AB magnitude $\sim$29 mag, similar to that of $z\sim 10$ galaxies in the Hubble Ultra Deep Field.

Throughout this paper, EW is referred to the total of \oiiidouble\ and \hb\ in the rest frame unless specified otherwise.
Magnitudes are calculated in the AB system. Errors are computed at $1\sigma$. 
We adopt a flat cosmology with $\Omega_{\lambda} = 0.7$, $\Omega_{M} = 0.3$ and $H_0 = 70\ {\rm km s^{-1} Mpc^{-1}}$.

\section{Data}
  \subsection{Data Reduction}

The CLASH data were obtained with three \HST\ cameras: ACS (Advanced Camera for Surveys)/WFC,
WFC3 (Wide Field Camera 3)/IR and WFC3/UVIS. 
All the 25 clusters have now been observed.
The images and catalogs are processed with {\tt APLUS} \citep{aplus}, 
which is an automatic pipeline modified from the {\tt APSIS} package \citep{blakeslee03} 
with an enhanced capability of processing WFC3 data and aligning them with the ACS data.
{\tt APLUS} processes the calibrated images from
the \HST\ instrument pipelines, namely the {\it flc} images for ACS (corrected for the detector's 
charge transfer efficiency) and {\it flt} images for WFC3/IR.
Recently, {\tt APLUS} has been updated so that images of individual exposure are aligned using \textit{DrizzlePac} \citep{astrodrz}, 
and the accuracy can achieve 1/5 pixel ($\sim 0 \arcsecpoint 015$) or better.

In the {\tt APLUS} process, images of different filter bands and cameras
were further aligned, resampled and combined
with a common pixel scale of 0.065\arcsec. 
We created detection images from the weighted sum of ACS/WFC and
WFC3/IR images and ran {\tt SExtractor} \citep{bertin96} in a
dual mode for all 16 bands. {\it mag\_iso} were chosen
in the color selections. We also verified the photometry
by comparing with public catalogs, \footnote{http://archive.stsci.edu/prepds/clash/}
which were processed with modified version of the \textit{Mosaicdrizzle} pipeline \citep{koekemoer03,koekemoer11},
 and no systematic deviation was found between the two pipelines.
In this paper, we focus on two ACS/WFC filters (F814W, F850lp) 
and five WFC3/IR filters (F105W, F110W, F125W, F140W, F160W),
which are hereafter called $I_{814}$, $Z_{850}$, $Y_{105}$, $YJ_{110}$, $J_{125}$, $JH_{140}$, and $H_{160}$ bands. 

\subsection{Sample Selection}
\label{sec:selection}
\begin{figure*}[]
\centerline{\includegraphics[scale=1.]{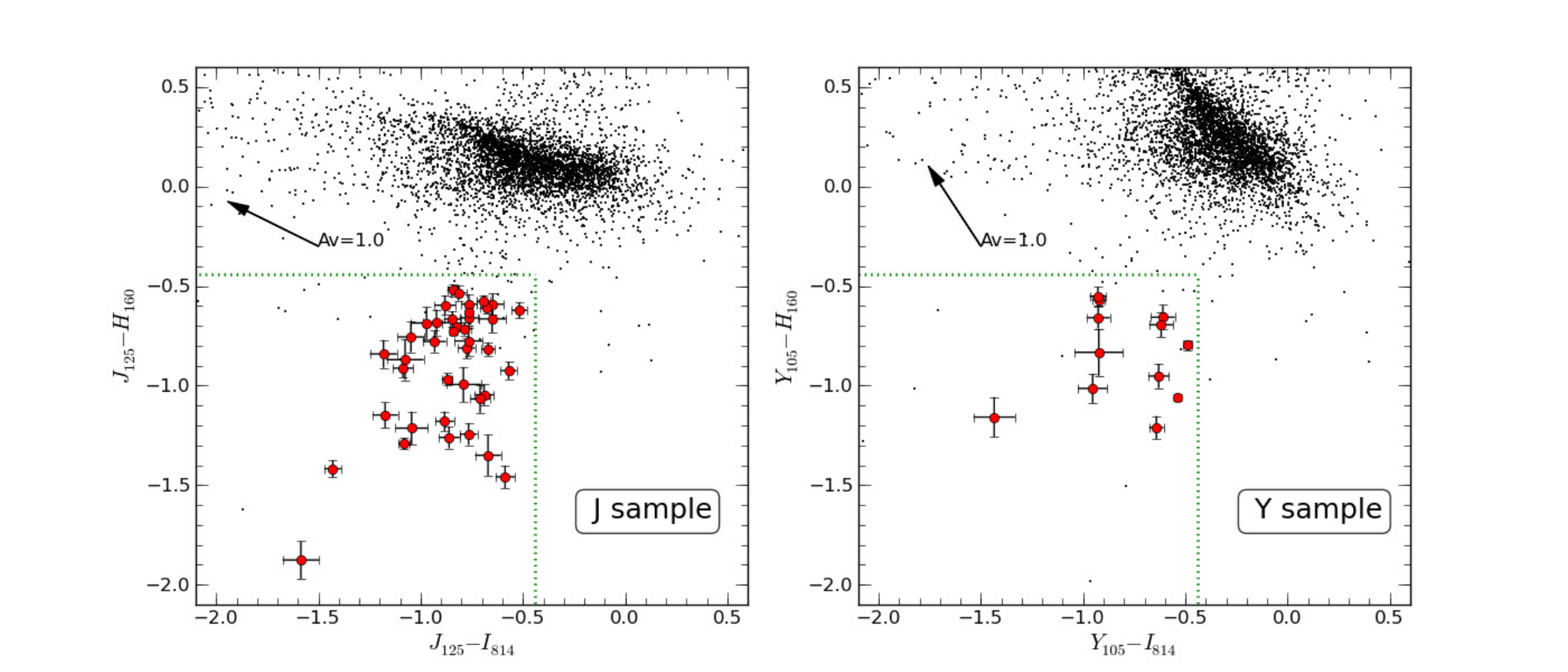}}
\caption{
Color-color diagrams to select the EELG candidates.
The red circles with error bars mark the selected EELGs 
and the black points mark all objects in the CLASH fields.
The regions separated by dash lines represent our selection criteria.
The black arrows illustrate the effect of dust attenuation using the formula from \citet{cardelli89}.
}
\label{color}
\end{figure*}

\begin{figure*}[]
\centerline{\includegraphics{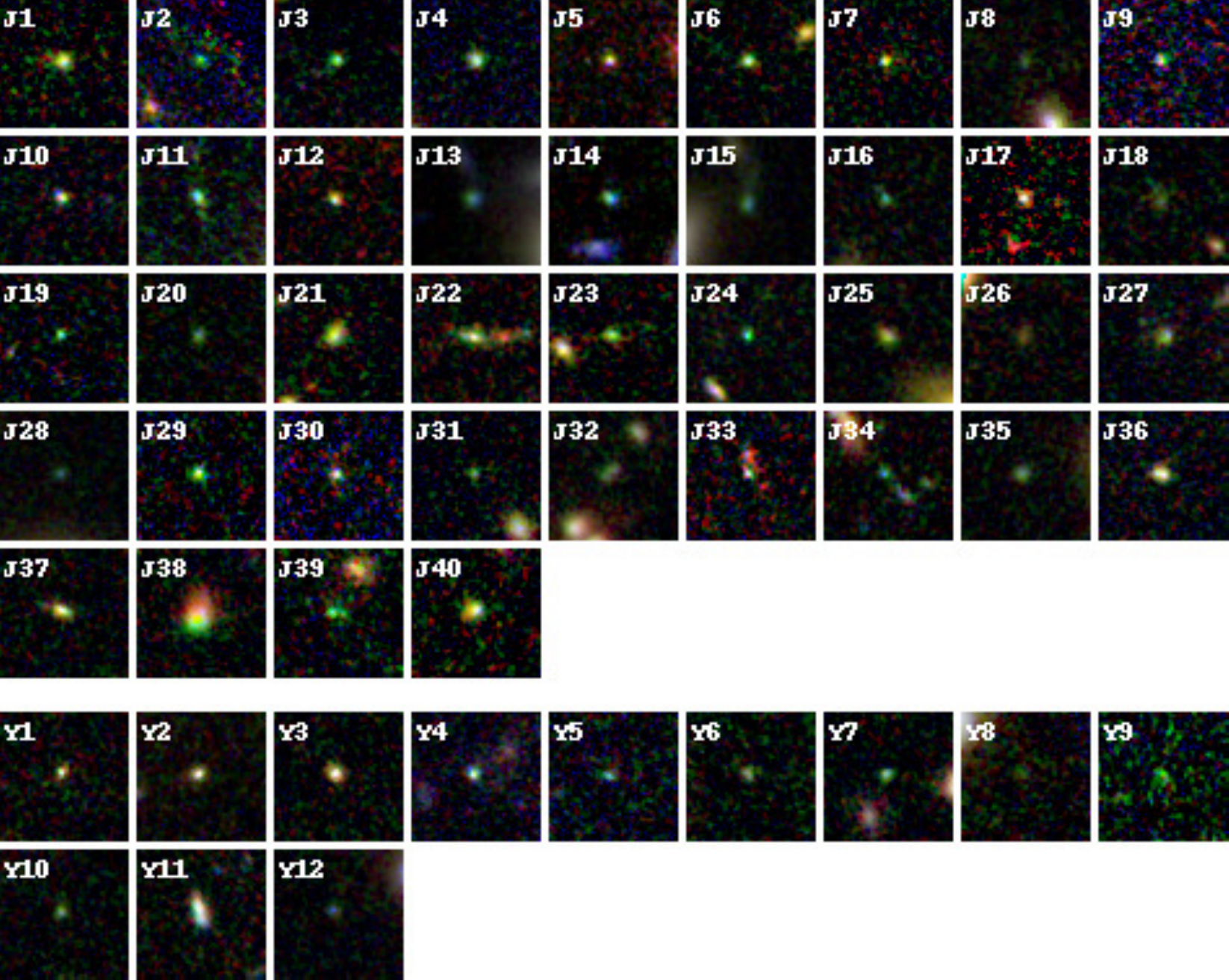}}
\caption{
Composite color images for the J and Y samples created with the same three bands as used for the selections.
The center peak bands are presented with green color, while the continuum bands are presented with blue and red colors.
The width of each stamp image is 50 pixels and the pixel size is 0.065"/pixel.
}
\label{rgb}
\end{figure*}

A color-color selection has been successfully used for identifying EELGs in VDW11.
We carried out two selections with two sets of filter bands.
Firstly, we follow the selection criteria of VDW11, namely
\begin{displaymath}
J_{125} - I_{814} < -0.44 - \sigma  \wedge
J_{125} - H_{160} < -0.44 - \sigma
\end{displaymath}
In the other selection, we use the $Y_{105}$ band as the central band, namely
\begin{displaymath} 
Y_{105} - I_{814} < -0.44 - \sigma  \wedge
Y_{105} - H_{160} < -0.44 - \sigma
\end{displaymath}
Where the $\sigma$ refers to the $1\sigma$ error of the color.
We also require that the three bands in each selection are detected at least $3\sigma$ to ensure good EW measurements. 
After these preliminary selections (Figure \ref{color}), 
we check the images and photometry of the WFC3/IR bands for each object.
Sources contaminated by cosmic-ray events, nearby bright sources and detector-edge 
effects are excluded.
We build two samples with 40 and 12 candidates named as the ``J'' sample and ``Y'' sample.
Note that the color excess of 0.44 magnitude in $J_{125}$ and $Y_{105}$ corresponds to rest-frame EW of about 600 \AA.
The false-color images of these galaxies are shown in Figure \ref{rgb}. 
We include the apparent angular sizes which are measured through full width at half maximum (FWHM) by {\tt SExtractor} in Table \ref{table1} and Table \ref{table2}.

\subsection{Redshift Estimation}
\begin{figure*}[]
\epsscale{1}
\includegraphics[scale=0.6]{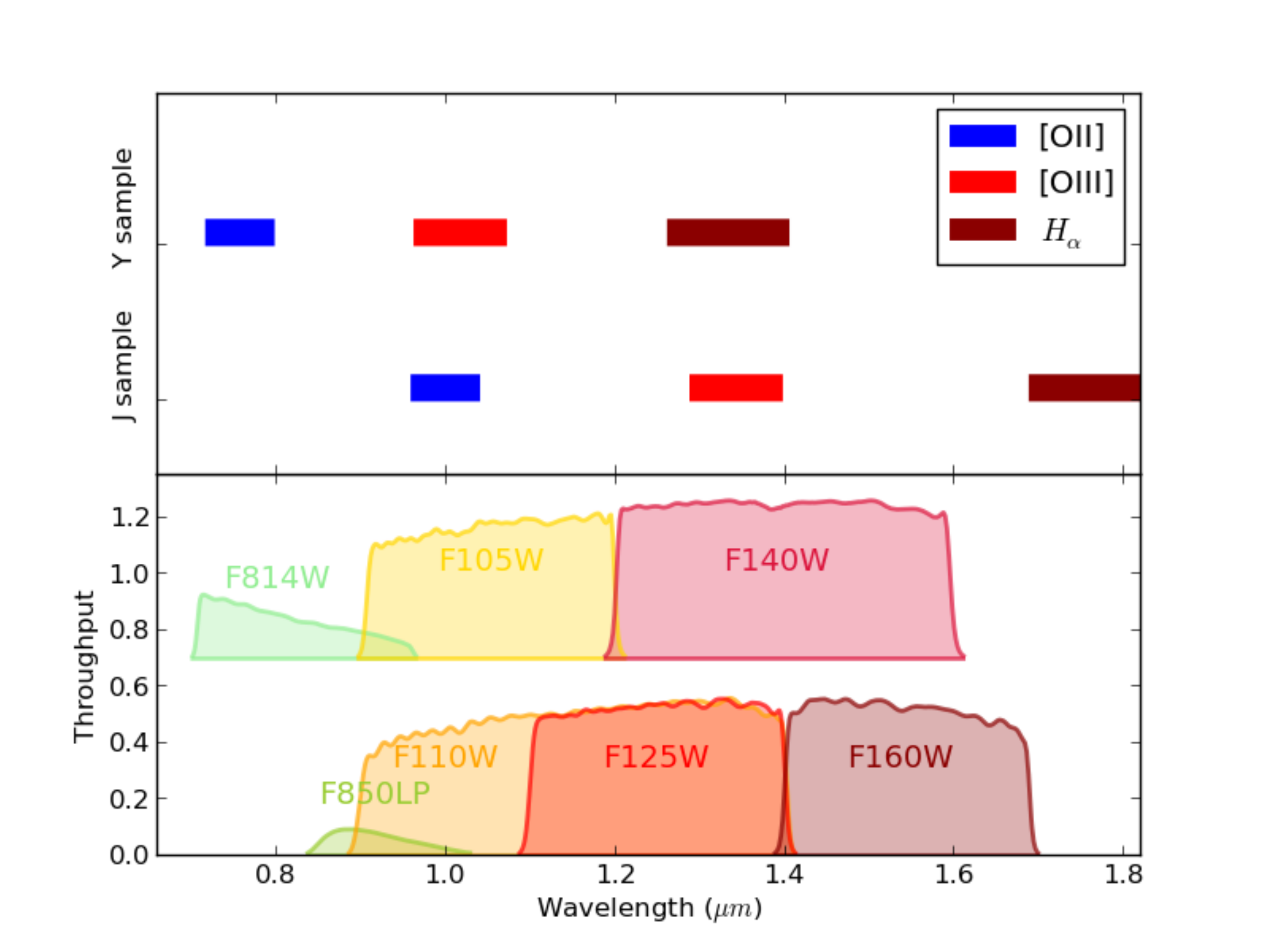}
\caption{
Wavelength ranges of strong emission lines in our J and Y samples.
The top panel shows the wavelength ranges for [OII], [OIII] and $H_{\alpha}$ with the redshifts of the J sample and the Y sample.
The lower panel shows the throughputs of the \HST\ filters. 
There is a total of 16 filters used in the CLASH observations,
but only the redward filters related to strong emission lines is shown here,  
}
\label{filter}
\end{figure*}

To illustrate the boosting effect in different bands and different redshifts, 
we simulate model spectra with a simple power-law continuum plus emission lines
and obtain the observed magnitudes using the throughputs of \HST\ filters (lower panel in Figure \ref{filter}).
The index of the power-law continuum is fixed at $\beta = 2$, which is defined 
as $F_{\lambda} \sim {\lambda}^{-\beta}$. Such continuum is a constant in different wavelength with AB magnitude system,
and is set to 28 mag. 
In the model, we choose metal-line lists from 
galaxies with sub-solar metallicity of $Z=0.2Z_{\sun}=0.004$ in \citet{anders03}
and only include emission lines with relative line intensities $F_{line}/F_{H_{\beta}}$ 
larger than 0.1.
The $H_{\alpha}$ line is included by assuming the ratio $H_{\alpha}$/$H_{\beta}$ = 2.86 from case B recombination \citep{storey95}.

\begin{figure*}[]
\epsscale{1}
\includegraphics[scale=0.6]{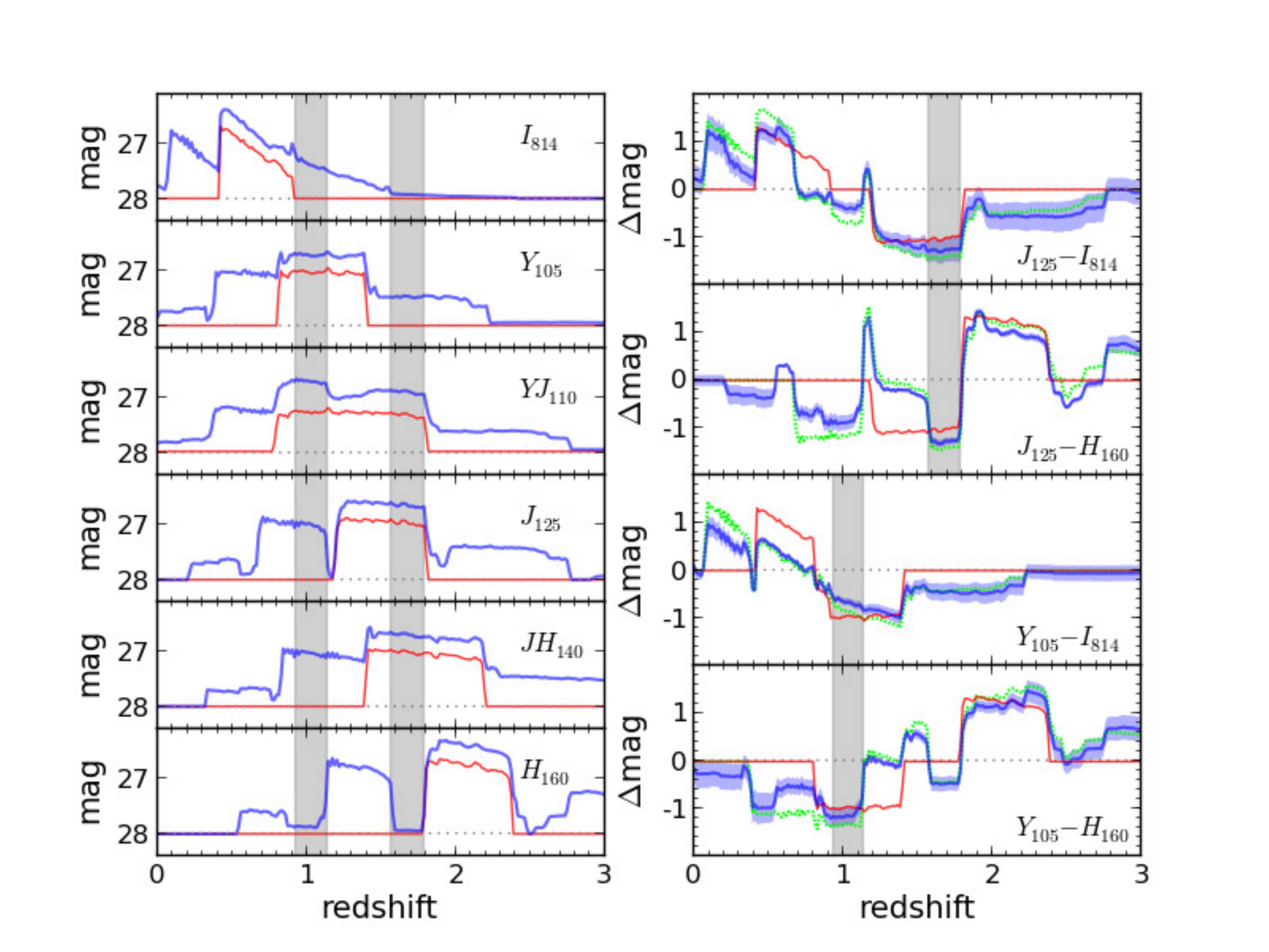}
\caption{
Effect of strong emission lines on the photometry as a function of redshift estimated with our spectral model.
The spectral model represents the spectrum of a star-forming galaxy, which consists of a power-law spectrum and strong emission lines with the flux ratios from \citet{anders03}.
In the left panels, photometry of six bands used in the EW calculations are plotted versus the redshift.
In the right panels, the colors used for our selections are plotted versus redshift (upper: the J sample, lower: the Y sample.).
In the spectral model, the two set of emission line ratios for metallicity $Z=0.2 Z_{\sun}$ and $Z= 0.02 Z_{\sun}$ are considered, which are shown with blue lines and green dotted lines.
The blue shadows around the blue lines in the right panels present the influence of spectral slope which changes from 1.5 to 2.5.
The contributions of \oiiidouble\ lines are shown with red lines.
Our two color-color selections are quite effective for selections of EELGs in two two redshift ranges ($0.93-1.14$ and $1.57-1.79$) marked with gray boxes.
}
\label{image}
\end{figure*}

In the model, the EW(\oiii) is set to 2000 \AA.
The simulated magnitudes and colors are shown with thick blue lines in Figure \ref{image}.
Based on this model, those EELGs in the redshift ranges $\sim 1.57 - 1.79$ (the J sample) and $\sim 0.93 - 1.14$ (the Y sample) are selected (grey regions in Figure \ref{image}).
The redshift ranges would not change significantly if we use a different EW in the model.
The upper panel of Figure \ref{filter} shows the wavelength ranges where the strongest emission lines, 
\ha, \oiii, \oii\ impact the observed flux.
For the Y sample, the \oii\ falls into the $I_{814}$ band, thus EELGs with stronger emission lines can be selected.

The spectral slopes of star-forming galaxies have been shown with a 1$\sigma$ dispersion of 0.4 \citep{bouwens09}.  
In the second model, we take into consideration the effect of spectral slopes 
and vary them between $\beta =1.5$ and 2.5.
The effect is shown in a blue shadow region in the right panel of Figure \ref{image}.
The changes in spectral slope would affect the color excess lower than 0.25 magnitude.
Some EELGs with blue slopes will not be selected. 
However, as the continuum is calculated with the average of the two bands at different wavelengths, 
the change of slope is not a problem in the following EW estimates.

Other emission lines, especially \ha\ and \oii, 
may also contribute to the broad-band photometry,  
but the relative flux to \oiii\ will vary due to the difference in metallicity, 
star formation history and extinction \citep{anders03,kewley04,salzer05}.
In order to show the boosting effect, we build a model with line intensities from 
metallicity $Z=0.02 Z_{\sun}$ galaxy (green dots in Figure \ref{image}).
In another model, we remove all other emission lines to show the contribution of \oiiidouble\ only (red lines in Figure \ref{image}).
While most emission lines do not affect the photometry as significantly as \oiiidouble,
\ha\ also boosts the magnitudes as its strength is similar to [\ion{O}{3}] in typical star-forming galaxies
and the \oii\ cannot be ignored.

As shown with the red lines in Figure \ref{image}, 
EELGs with redshift 1.14--1.57 would also be selected into the Y sample or J sample, only if \ha\ is extremely weak.
The \oiii\ lines can be excited by massive stars as well as active galactic nuclei (AGN).
The relatively strong \oiii\ and weak \ha\ can be due to either metal-poor star-forming galaxies or the contribution of AGN \citep{kauffmann03}. 
To exclude these EELGs from the J sample is not possible with current observations.
In the Y sample, these EELGs can be identifiable by comparing the magnitudes of $JH_{140}$ and $H_{160}$.
As the \ha\ falls in both the two bands in these galaxies, the two bands should be observed with similar magnitudes as shown in the left of Figure \ref{image}.
We find 5/12 candidates in the Y sample have similar $JH_{140}$ and $H_{160}$ magnitudes.
If parts of these weak $H_{\alpha}$ EELGs with low $H_{\alpha}/[OIII]$ ratios are AGN, 
the upper limit of the AGN fraction in our Y sample is about $42\%$ (5/12),
which is still consistent with the AGN fraction ($\sim17\%$) estimated with a larger spectral emission line galaxy sample in \citet{atek11}.
As our analysis is not sensitive to the redshift, 
we assume that the redshift ranges are $z \sim 1.57 - 1.79$ and $0.93 - 1.14$ for the J and Y samples, respectively.

\subsection{Magnification}
The major advantage of CLASH observations is that cluster lensing 
provides a powerful tool to enable us to discover intrinsically faint galaxies.
The magnification maps for all 25 clusters are made based on the strong lensing model of \citet{zitrin09,zitrin11}.
As magnification factors are redshift dependent, 
we use the median redshift 1.03 for the Y sample and 1.68 for the J samples in calculations.
However, the following EW calculations are based on the color excesses, and thus independent of magnifications.
The uncertainties of magnifications are lower than 10\% for the ranges of redshifts. 
The estimated magnifications are likely in consistent with the true value at 68\% confidence for magnifications lower than 5 \citep[Figure 11 in ][]{bradley13}. 
The uncertainties are only significant for those high magnified candidates.

\begin{figure*}[]
\centerline{\includegraphics[scale=0.6]{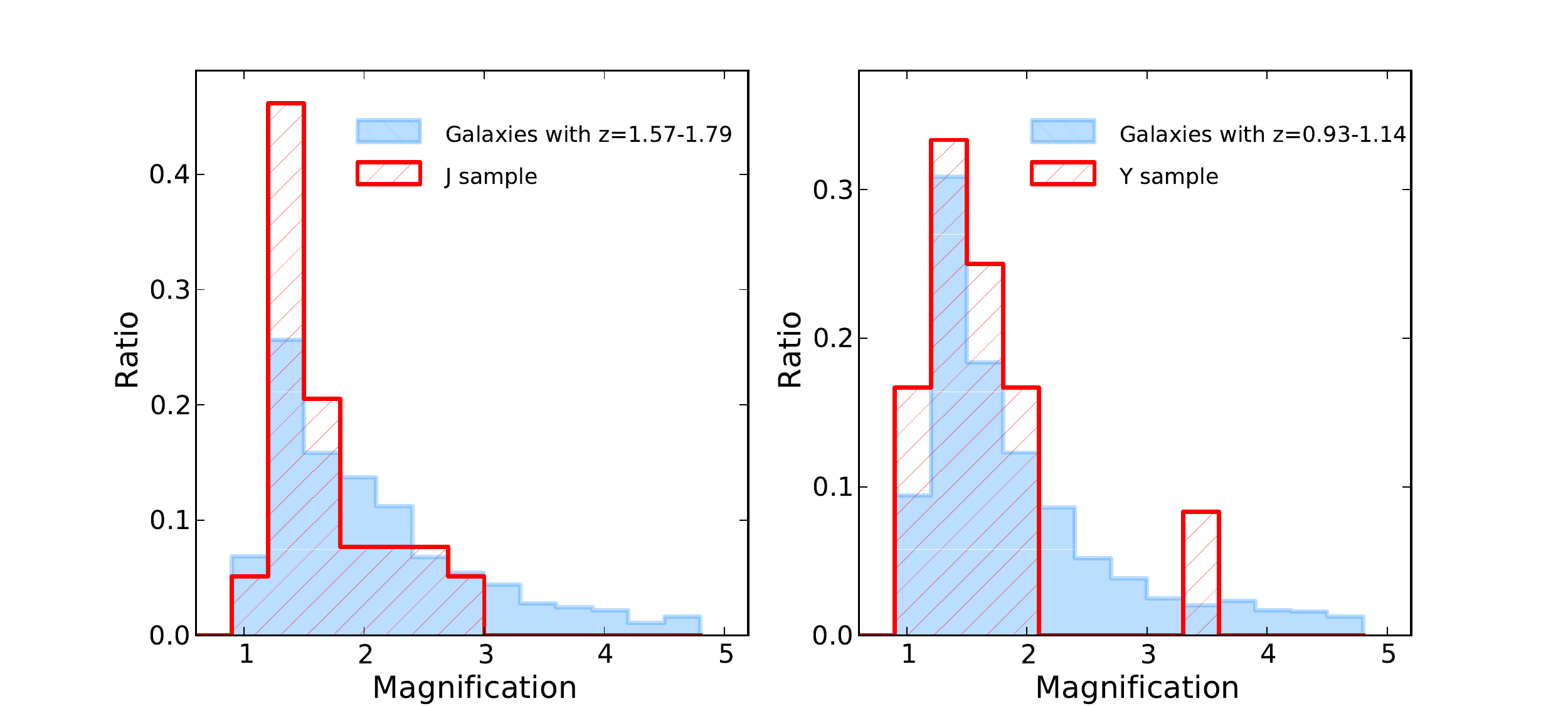}}
\caption{Magnification distributions of our J and Y samples (red histograms)
and galaxies in the same redshift ranges from the cluster fields (blue filled histograms).
}
\label{mag}
\end{figure*}

The source magnification factors are listed in Table \ref{table1} and \ref{table2}.
Candidate J22 has a magnification as large as 8.4 and shows an apparent extended structure.
The delensed magnitude is estimated as $\sim 28$ mag.
Its EW is higher than 2000 \AA, as confirmed with the color excess in $YJ_{110}$, $J_{125}$, and $JH_{140}$.
The magnification distributions of our two samples are shown in Figure \ref{mag} in red color.
We also calculate the magnifications for field galaxies that are within the same two redshift ranges 
as our samples in all 25 clusters, 
and show the distributions in filled blue histograms.

\section{EW Estimates}
\label{sec:ew}
Due to the lack of deep near-infrared spectroscopic data for our samples, 
we estimate the EWs by comparing the flux excess between
the bands boosted by the strong emission lines (the central peak bands)
and the adjacent bands dominated by the continuum (the continuum bands).
As shown in Figure \ref{filter} and \ref{image}, the strong emission lines boost more than two bands,
thus the combination of EWs estimated from different boosted bands makes the EW more reliable.

\subsection{Method}
\label{sec:method}
The EW is estimated using two continuum bands and one central peak band through:
\begin{equation} 
  \rm{EW} =  \frac{F_{total}-F_{c}}{F_{c}}
  \frac{{\rm{W}}}{\it{1+z}} 
\end{equation}
where $F_{total}$ is the flux of the central peak band which is the total flux of emission lines and continuum,
and $F_c$ is the continuum flux in the $I_{814}$ and $H_{160}$ bands. 
$W$ is the effective width of the central peak bands.
$z$ is the redshift of the EELGs and is set to the median redshifts, 
which are 1.03 and 1.68 for the Y and J samples, to translate EW to the rest frame.
An accurate continuum analysis is important to estimate the EW.
In the procedure, we assume the spectral slopes equal 2 to estimate the continuum based on the $I_{814}$ and $H_{160}$ bands,
and use the weighted averages as the continuum at the central peak bands.
In the redshifts of our two samples, the rest frame UV continuum are observed by ACS/WFC bands,
thus the UV slopes can be obtained by linear fits of these bands
(Column 11 in Table \ref{table1} and Table \ref{table2}).
The average slopes of the J and Y samples are  $2.06\pm0.02$ and $2.07\pm0.03$, 
consistent with our assumption and other studies of strong emission line galaxies \citep{van11, van13}.

The uncertainties in the EW measurement is because that other emission lines
could affect the $I_{814}$ and $H_{160}$ photometry and cause an overestimate of the continuum level
and subsequently an underestimate of the EW.
Within our model of $EW = 2000 \AA$,  the continuum is boosted by less than 0.1 mag for the J sample but 
larger for the Y sample, in which the \oii\ line falls into the $I_{814}$ bands.
Therefore, the EW should be considered as a lower limit in this situation.
Furthermore, the continuum bands are considerably fainter than the center peak bands.
Even if the continuum flux is estimated by averaging two continuum bands,
the continuum errors are still the main affecting factors of the EW measurements.
For some faint EELGs, their continuum flux cannot be well constrained by photometric data due to these facts,
which limit the accuracy of EW calculations.

\begin{figure*}[]
\centerline{\includegraphics[scale=0.8]{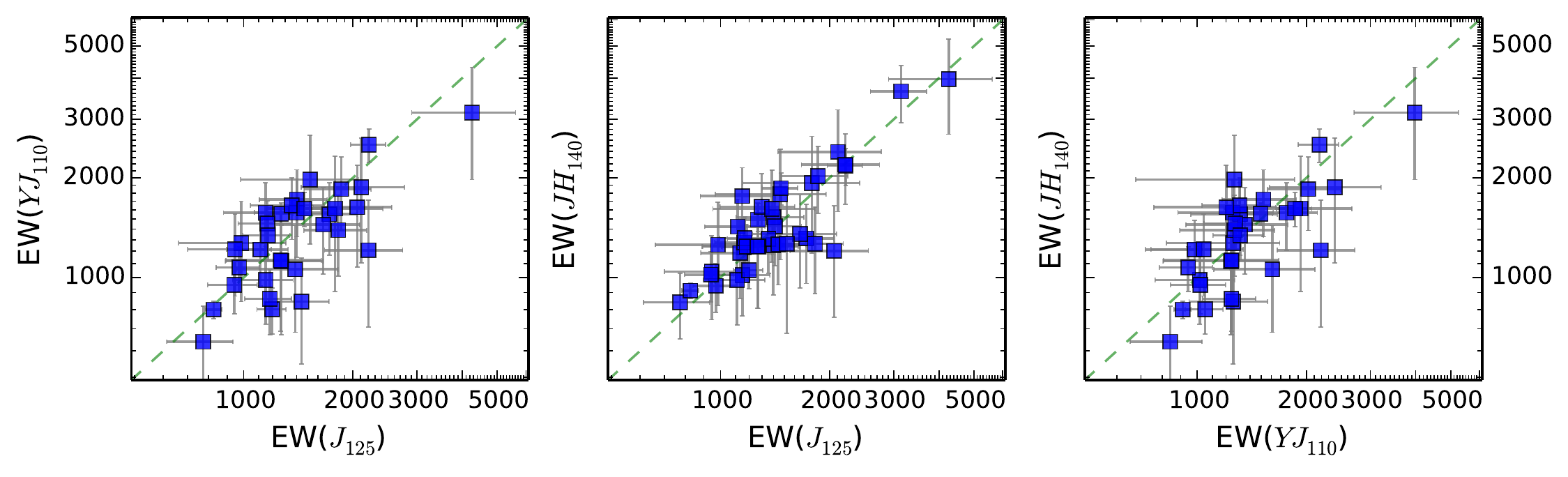}}
\caption{Comparison of EWs derived from the color excesses in three different bands for the J sample with the method described in section \ref{sec:method}.
   In these calculations, the continua are derived from $Y_{814}$ and $H_{160}$, 
   but the EWs are derived from the excesses of $YJ_{110}$, $J_{125}$ and $JH_{140}$, respectively.
   The widths of $YJ_{110}$ and $JH_{140}$ are broader than $J_{125}$, thus the EWs derived from
   these two bands involve larger uncertainties.
   The consistency between EWs suggests the robustness of the EW measurements.
   }
\label{ewcompare}
\end{figure*}

As shown in Figure \ref{filter} and \ref{image},
all the $YJ_{110}$, $J_{125}$, $JH_{140}$ are covered by the [OIII]+$H_{\beta}$ for the J sample,
thus the EWs can be constrained from the color excesses of these three bands.
Then the measurements can be verified by comparing EWs calculated from different center peak bands (Figure \ref{ewcompare}).
The uncertainties in EW measurements are slightly higher for broader bands such as $YJ_{110}$.
Nonetheless, the consistency of EWs inferred from three broad bands makes our results robust.

At the redshift of the Y sample (Figure \ref{image}), 
\ha\ has moved into the wavelength ranges of $J_{125}$ and $JH_{140}$,
both [OIII]+$H_{\beta}$ and \ha\ are covered by the $YJ_{110}$.
The $Y_{105}$ band is boosted by [OIII]+$H_{\beta}$ only.
The EWs derived from the color excess are included in Table \ref{table2}.

\subsection{EELGs with Extremely High EW}

\begin{figure*}[]
\includegraphics[scale=0.5]{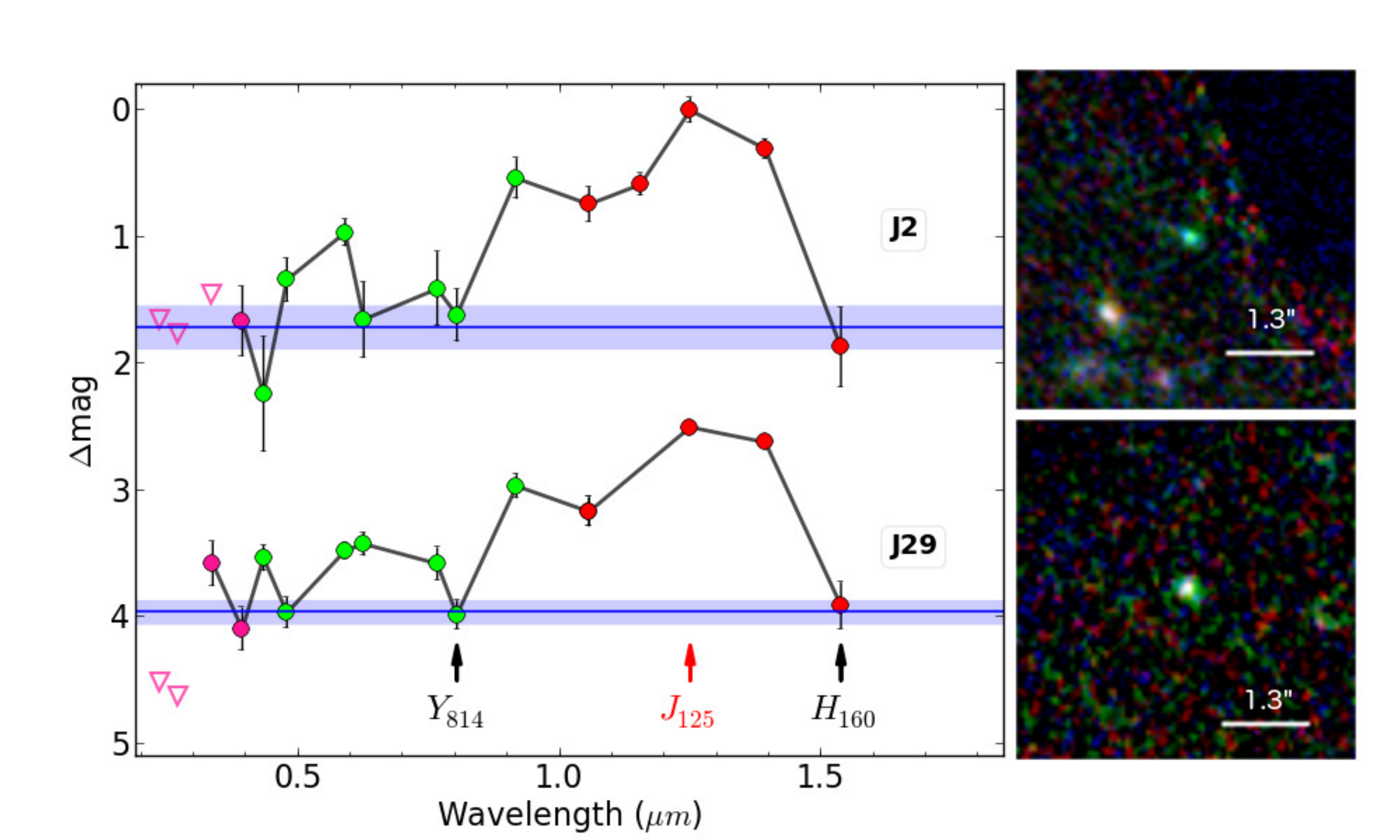}
\caption{
  SED for the two EELGs with ${\rm EW} \sim 3000$ \AA\ in the J sample (pink: WFC3/UVIS bands, green: ACS, and red: WFC3/IR). 
  The black and red arrows show the central wavelengths of $I_{814}$, $J_{125}$ and $H_{160}$ that are used in our selection.
  The solid blue lines are the estimated continuum levels based on the $I_{814}$ and $H_{160}$, and the shaded regions show the $1 \sigma$ error.
  The source fluxes are remarkably boosted by strong emission lines up to 1.5 magnitude.
  The composite color images in the right are created with $I_{814}$ and $Y_{105}$ (blue),
  $YJ_{110}$, $J_{125}$ and $JH_{140}$ (green), $H_{160}$ (red).
}
\label{highew}
\end{figure*}

In the J sample, the magnitudes in $Y_{105}$, $YJ_{110}$ and $JH_{140}$ are all boosted by \oiiidouble, 
and the derived EWs can be averaged with the errors as weights to make a better constraint.
There are two EELGs (J2 and J29) among the 40 EELGs showing EWs above 3000 \AA,
which are $3737\pm726$\AA and $3332\pm439$\AA.
As seen in Figure \ref{highew}, both of  these two EELGs are extremely compact.
The FWHM of the instrument point-spread function (PSF) is $0 \arcsecpoint 14$ for the $J_{125}$.
hence the EELGs are all resolved.
After subtracting the instrument PSF and the lensing effect, 
the measured sizes correspond to physical sizes of about 1.7 kpc and 0.9 kpc.

In addition, the EW of Y9 in the Y sample is also measured to be $3307\pm1293$ \AA.
As only the $Y_{105}$ is free from contaminations by other emission lines in the Y sample, 
the EW can be only derived from the excess of the $Y_{105}$ band and thus contains higher uncertainty. 
Y9 is also selected as a LBG in \citet{bradley13} (B13 hereafter) and we will discuss this candidates in the next section.

\begin{figure*}[]
\includegraphics[scale=0.8]{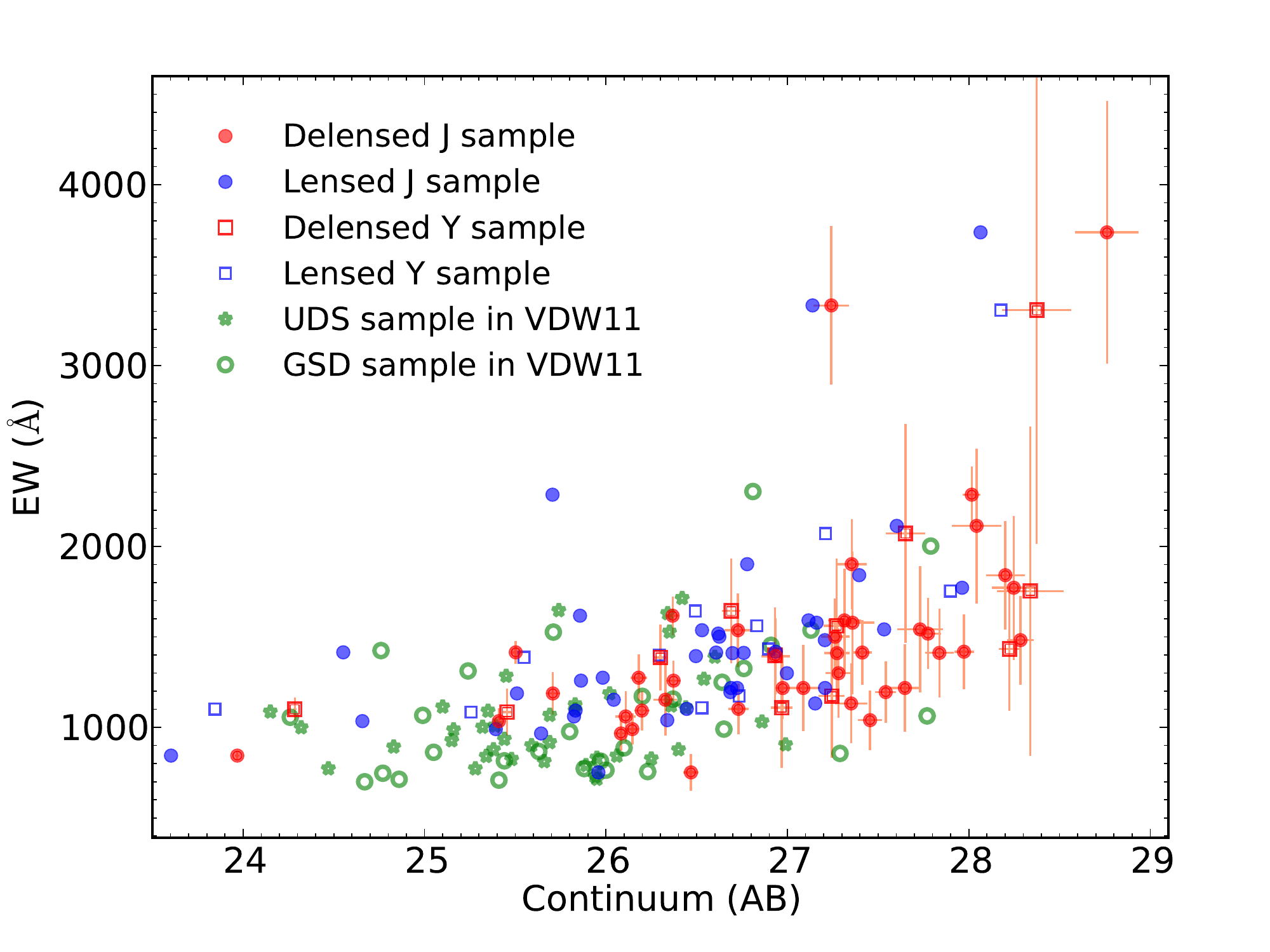}
\caption{
EW vs. continuum luminosity for EELGs in the J and Y samples and the UDS and GSD samples from VDW11.
The J and Y samples are marked with filled circles and open squares,
in which the observed magnitudes are shown in blue color and
the intrinsic (delensed) values are shown in red color.
As a comparison, the UDS and GSD samples are plotted with green stars and circles.
EWs in the J sample are the weighted averages of the three EWs calculated from three different bands.
EWs in the Y sample are calculated with the excesses in the $Y_{105}$ only.
}
\label{ewvsl}
\end{figure*}

The same selection method as our J sample is also utilized for CANDELS field in VDW11.
VDW11 identified 69 candidates in a total area of 279 square arcminutes.
The typical area coverage over each cluster field in CLASH is $\sim 4\ arcmin^2$ 
indicated from the area available within the WFC3/IR field of view.
To count for the influence of the bright galaxies in the fields, 
we count the areas covered by galaxies based on the segmentation images and subtracted these areas
from the total areas indicated from exposure time images.
Then the total areas to search EELGs in CLASH fields is $\sim 112\ arcmin^2$.
The VDW11 used the data from the Ultra Deep Survey (UDS) field in the wide program,
and the GOODS-South Deep (GSD) field at 4-epoch depth in the deep program.
The VDW11 sample include 40 and 29 EELGs from these two fields, named as the UDS sample and GSD sample hereafter.
The number density for the J sample is 0.36 per square arcminutes, 
compared to 0.19 per square arcminutes in the UDS sample and 0.39 per square arcminutes in the GSD sample. 
The high detection rates in the J sample and the GSD sample are due to the deeper observations.
As shown in Figure \ref{ewvsl}, the J sample is the deepest and can reach about 28.5 mag after correcting the lensing effect.
The lensing effect increases the depth and reduces the volume of observations in the meantime,
in addition of the small samples,
we do not find significant different detection rate between the CLASH fields and the GSD field.

It is apparent in Figure \ref{ewvsl} that the EELGs with higher EWs are more common in fainter EELGs.
The three EELGs with extremely high EWs are fainter than most candidates. 
In the VDW sample, only two candidates has EWs larger than 2000 \AA, 
which are $2304\pm515$ \AA\ and $2002\pm849$ \AA
\footnote{VDW11 used the EW(\oiii) instead of EW([OIII]+$H_{\beta}$) in their table.
The EW(\oiii) were calculated by assuming a fixed flux ratio of $H_{\beta}$ and \oiiidouble.
Therefore, we obtain the EW([OIII]+$H_{\beta}$) using the same ratios.}. 
Both of these two candidates are selected from the GSD field which are deeper than the UDS field.
All these samples support the idea that we can detect EELGs with stronger emission lines in deeper observations.
The three extreme candidates are unlikely due to noise fluctuations.
We examine the probability that one EELG with EW([OIII]+\hb)=2000 \AA\ at the same redshift 
are measured to be $EW \ge 3000 \AA $. 
In our spectral model, fluctuations with the level of 1/3 continua, which are the worst cases in our samples,
are added to each band.
We simulate the photometry for 10000 times and measure the EWs with the same method.
The possibility of spurious large EW ($\ge 3000 \AA$) is lower than 20\%.
The rate will reduce to lower than 4\%, if the EW are measured with flux excesses in two bands,
and the rate will be negligible if there are three bands to constrain the EW.
As the spectra in Figure \ref{highew} and the results in Table \ref{table1}, 
the large EWs are confirmed with three bands for J2 candidate and two bands for J29 candidate.
Both the two EELG candidates show similar spectral shapes,
moreover, the significant boosts in $Z_{850}$ and $Y_{105}$ due to \oii\ line are also in agreement with the strong [O III] 5007\AA\ line.
The EW of Y9 candidate can only be estimate with the boost of $Y_{105}$, 
therefore, the noise fluctuation cannot be totally excluded and the strength could be confirmed with further observations.

VDW11 found these candidates are low-mass ($\sim 10^8 M_{\sun}$) galaxies, starbursting ($\sim 5 M_{\sun}\ yr^{-1}$),
young ($5\sim 40\ Myr$) galaxies.
The nature is also consistent with the spectral observations of two highly lensed EELGs at z=1.85 and 3.12 \citep[][respectively]{brammer12, van13}.
We estimate the age and stellar mass following the same method as VDW11.
In general, we use the Starburst99 model \citep[SB99,][]{leitherer99} with continuous star formation 
and a \citet{chabrier03} IMF with a high-mass cut off at 100$M_{\sun}$ and metallicity 0.2$Z_{\sun}$.
The EW($H_{\beta}$) declines as time and is used as an age indicator. 
EW($H_{\beta}$) is assumed to contribute 1/8 to the combined EW.
The upper limit of EW([OIII]+$H_{\beta}$) is 4336 \AA\ from this model
and will decrease to 3838 \AA\ if we use a solar metallicity.
VDW11 estimated stellar masses based on the rest frame V band luminosity from the $H_{160}$ band photometry.
To increase the accuracy, we use the weighted average photometry of the continue bands instead. 
The inferred masses are shown in Table \ref{table1} and Table \ref{table2}.
The median value of $7.1\times 10^6\ M_{\sun }$ is one order lower than the VDW sample. 
\citet{maseda13} find similar dynamic mass and the young stellar mass ratios of the VDW sample to other star forming galaxies in $z\sim2$,
and confirmed that the low stellar mass are dominated by the intense starbursts.
These EELGs provide insight into the evolution of the dwarf galaxies and
provide evidence that the starburst phase plays a key role in the mass build-up for at least some low mass galaxies.

\section{Discussion}

Strong emission lines are known to have a significant impact on the SED fitting
\citep{atek11,schaerer09,labbe10,labbe12,stark13,shim11,fumagalli12}.
\citet{labbe12} investigated the ultra-deep Spitzer/IRAC photometry for a sample of $z \sim 8$ galaxies from 
the Hubble UDF program and found an average contribution of $\sim 0.44$ mag to the [4.5] band of Spitzer by \oiiidouble.
\citet{schaerer09} found that the apparent Balmer breaks  
observed in a number of $z \sim  6$ galaxies detected at $>3.6$ \micron\ with
Spitzer/IRAC can be explained by the presence of redshifted strong emission lines.
\citet{smit14} select galaxies at narrow redshift range $z \sim 6.6 - 7.0$ to avoid contamination of other emission lines
and detected very high $[OIII]+H_{\beta}$ lines.
The mean value of $EW([OIII]+H_{\beta})$ derived from the excesses of 3.6 \micron\ band is greater than 637 \AA\ with one extreme EW of 1582 \AA.
\citet{shim11} also found that the fraction of emission line galaxies evolves with redshift and that 
 emission-line galaxies could be more common in high redshifts. 
However, properly considering the impact is still a challenge due to the lack knowledge of such galaxies.
The EELG samples are the median redshift analogs and provide a good opportunity to study the high redshift star forming galaxies.

While emission lines have been considered in galaxy templates \citep{schaerer09, ono10},
the high redshift galaxy selection itself can also be affected due to the exist of median reshift EELGs. 
The fainter and stronger EELGs in our samples indicate that the impact of EELGs to the selection of LBG has probably been underestimated.
The properties of LBG could be misunderstood due to the mix of these EELGs.
\citet{taniguchi10} investigated the EELG interlopers
for $z\sim$8 galaxies in the Hubble Ultra Deep Field. 
The EWs of \oiii\ in their models only vary up to $\sim$1000 \AA, 
therefore they claimed a negligible probability for low-redshift interlopers.
Considering the EELGs with EW([OIII]+$H_{\beta}$) $\geq$ 3000\AA\ in our sample
and the strong \oiii\ and \ha\ in other surveys 
\citep{cardamone09,atek11, brammer12, shim13, van13, smit14},
it is necessary to reexamine the impact of EELGs on the selections of LBG.

\subsection{Contamination to the Selection of $z \sim 6$ Galaxies} 

Firstly, we test whether any sources in our samples would satisfy the color selections for LBG.
From Figure \ref{color},
both the terms of $J_{125}-I_{814}$ in the J sample and $Y_{105}-I_{814}$ in the Y sample
can reach more than one magnitude.  
In the color-color selections of LBG,
a decrement of one magnitude is adopted \citep{bouwens12b, zheng12, oesch12, oesch13}.
Furthermore, a candidate at $z>7$ must not be detected in the optical bands. 
For the bright EELGs in our samples, their continuum from near infrared to UV bands is detected with high confidence.
However, there are still a few EELGs with faint continua and the bands bluer than $I_{814}$ fall bellow the detection limit,
 resulting in mimic LBG with redshift around 6.
When the S/N ratio is low, 
it becomes difficult to distinguish whether the color excess is due to the Lyman break as seen in LBG 
or the boost by strong emission lines in EELGs. 
 
\begin{figure*}[]
\centerline{\includegraphics[scale=0.5]{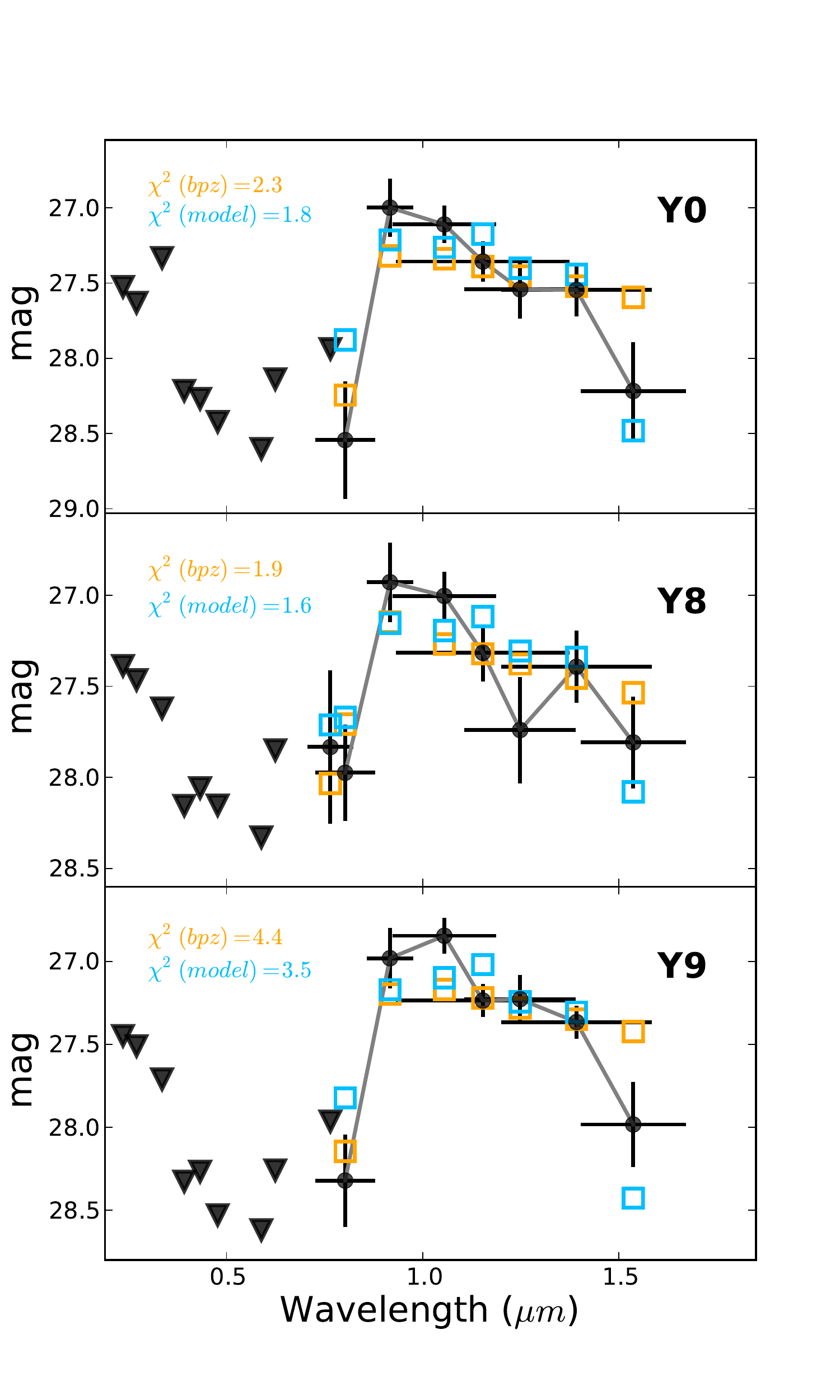}}
\caption{
Observed magnitudes and best-fit SED of three EELG candidates that can also be selected out as high redshift LBGs.
The observed photometries are shown with black circles and the triangles represent the $2\sigma$ detection limits.
All four objects display color excesses near the $Y_{105}$
which can be fit with both LBG model(orange open squares) and the EELG model (blue open squares).
In the EELG model, the spectral slope is fixed to 2 and EW([OIII]+$H_{\beta}$) is fixed to the value estimated from color excesses in $Y_{105}$.
Other metal emission lines with the flux ratios of stellar metallicity with 0.2$Z_{\sun}$\citep{anders03} have been added.
The only two free parameters in the models are the redshift and normalization factor. 
 }
\label{highz}
\end{figure*}

Recently, B13 reported a considerable number of galaxy candidates at $z\sim 6-8$
in 18 CLASH clusters
(Abell 1423, Abell 209, CLJ1226.9+3332,  MACS0429.6--0253, MACS1311.0--0310, MACS1423.8+2404, RXJ2129.7+0005 are not included compared to the total of 25 clusters).
Their selections are based on the redshifts calculated from Bayesian photometric redshift (BPZ) code \citep{benitez00}.
This method also identifies high redshift galaxy candidates primarily based on the Lyman break feature
and the results are generally in very good agreement with the common color-color selection method.
Possible contaminations of EELGs at the high redshift samples are shown by matching the samples to our EELGs.
Two EELGs, Y8 and Y9,
are also selected as m1115--0352 (z=6.2) and m1720--1114 (z=5.9) in the z$\sim$6 sample of B13.
Another galaxy in Abell 209 (RA: 22.954264, DEC: -13.611176), 
which is removed out from the Y sample because the $I_{814}$ band detection is below $3\sigma$,
is also included for its high photometric redshift (5.89) from BPZ and is called Y0.
These candidates have similar spectra shapes as shown with solid black circles in Figure \ref{highz}.
The best BPZ results are shown with open orange boxes.
We also fit the $I_{814}$ band and redward bands with our strong emission line model which is shown with blue boxes.
In the emission line model, we fix the spectral slope to 2 and the EW([OIII]+$H_{\beta}$) to be the value estimated with $Y_{105}$ band excess.
Emission lines with flux ratios from a 0.2$Z_{\sun}$ galaxy \citep{anders03} are considered.
The only two free parameters left are the redshift and the normalization.
While the high redshift assumption often fails to fit the $H_{160}$ band, 
our emiison line model can explain the drop of flux in $H_{160}$ bands in three galaxies.
The emission line model also overestimate the flux in the $Y_{814}$ band.
The ${\chi}^2$ values are shown with the same colors in the figure. 
For all the three candidates, the spectra favor the emission line models for the slightly lower ${\chi}^2$ values.

If such candidates are chosen as LBG, 
the UV-continuum slopes derived from infrared bands would be misleading.
We estimate the slopes for the three candidates with a linear fitting method for the bands redder than the $I_{814}$ band.
The derived UV slopes are $3.7\pm0.5$, $3.6\pm0.5$ and $3.5\pm0.4$ for Y0, Y8 and Y9, respectively. 
\citet{bouwens10} obtain the UV-continuum slopes for redshift 6 galaxies with $Y_{105}$, $J_{125}$ and $H_{160}$.
Using the same method, we get even bluer slopes which are $4.5\pm0.8$, $4.9\pm0.9$ and $4.4\pm0.6$.
These values are all at least $1\sigma$ bluer than the mean UV-continuum slope in redshift 6 \citep{bouwens10}.
Nevertheless, the current CLASH data and Spitzer/IRAC data are not deep enough to confirm the nature of these galaxies.
Our EELG color--color selections are limited by the luminosity of continuum.
Much more EELGs with stronger emission lines are faint and below our detection limit according to the EW and luminosity trend.
What is more, our samples only include EELGs in specific redshift ranges.
Therefore, the total contamination of EELGs to the high redshift sample of B13 can be higher.

  \subsection{Impact on the Selection of $z \sim\ 10$ Galaxies}
Recently, the search for LBG has reached $z \sim 10$ \citep{zheng12, bouwens13a, ellis13, coe13, oesch13}.
One of them, UDFj-39546284, has already been found to be possibly an EELG at redshift $\sim$2 \citep{brammer13, capak13}.
More candidates are found in the ongoing Hubble Frontier Fields \citep{zheng14, zitrin14, ishigaki14}.
As there is no enough spectral observations to confirm the redshifts,
properly considering the contamination of EELGs are quite necessary for these samples. 

For galaxies with $z\sim8.7-10.5$, the \lya-break shifts into the $JH_{140}$ band and could be selected as $J_{125}$ drop-out.
If the redshift is higher than 10.5, the \lya-break has shifted out of the $J_{125}$ band, and
only $H_{160}$ and parts of $JH_{140}$ cover the continuum.
EELGs with strong emission lines falling in the $H_{160}$ band can also mimic such spectral feature.
Considering a power-law spectrum with EW([OIII]+$H_{\beta}$) = 2000(3000) \AA\ in our model, 
the dropout $J_{125}-H_{160} \sim 1.1(1.4)$ would be observed for EELGs at redshift 1.9.
Taking into account the deviation of the slopes which would increase the observed color difference in EELGs, 
we propose that, 
unlike current HST WFC3/IR surveys in CLASH or UDF12,
at least one mag deeper observations in the bluer bands will be required to constrain the flux of the continuum and 
draw a distinction between these two scenarios. 
In EELGs at the matched redshift, the \ha\ has moved out of the $H_{160}$ band, but the impact of \oii\ on $J_{125}$ cannot be neglected,
therefore the deeper observations in the $Y_{105}$ or $YJ_{110}$ bands will be helpful.
If a deeper observation is not available, a significant detection in Spitzer observation are required to confirm the high redshift LBG \citep{zheng12, oesch13}.
Real high redshift LBG have a flat UV continuum and 
the detection from $H_{160}$ band to mid-infrared bands cannot be mimicked by any type of EELGs.

 \subsection{EELG Fraction in High Redshift Samples}
 
While it is almost impossible to distinguish EELGs and high redshift galaxies without spectroscopic observations,
we should consider that there is a certain number of EELGs in the high redshift galaxies.
The spectral features of our J and Y samples are similar to the $I_{814}$ and $Y_{105}$ dropout galaxies at redshift range 5.6-8.0.
The B13 sample includes a total of 206 galaxies in this redshift range 
and we find two galaxies common as our EELG samples.
If the both of these two galaxies are EELGs, 
the contamination fraction for the B13 sample is at least $2/206\sim1\%$.
However, the real contamination fraction could be higher due to the limit of observations.
Our EELG samples only include such candidates that the continua are detected higher than $3\sigma$ in $I_{814}$.
There are substantial fainter EELGs which probably include stronger emission lines ( See the trend in Figure \ref{ewvsl}).

Rencently, the frontier of high redshift samples are redshift higher than 9 selected based on the $J_{125}$ drop-out.
The fraction of EELGs in these $J_{125}$ drop-out samples will be more significant than the contamination in B13 sample.
Firstly, these $J_{125}$ drop-out samples are in redshift range 8.7-10.5 and can be contaminated by EELGs in redshift range 1.8-2.2.
The comoving volume of the redshift range 1.8-2.2 is 1.1 times of the total volume of the Y and J samples.
Meanwhile, the comoving volume of the $J_{125}$ drop-out samples is 0.6 times of the $I_{814}$ drop-out samples.
In the same projected area, there will be $1.1/0.6\sim1.8$ times more contaminations due to EELGs.
Secondly, the buildup of star-forming galaxies peak at near redshift 2 
which means there are a substantial population of EELGs in that redshift.
From the median redshift of our EELG samples 1.5 to redshift 1.9,
the space density of star forming galaxies increases by 2.3 times \citep{oesch10}.
Moreover, the space densities of star-forming galaxies decline since redshift around 2 \citep{bouwens14}. 
From redshift 6 to redshift 9, the space density of galaxies decreases about ten times.
All of these will result in that there are the EELG contamination fraction for the $J_{125}$ drop-out samples 
will be $1.8\times 2.3\times 10\approx 41$ times higher.
\citet{pirzkal13} develops a Markov Chain Monte Carlo (MCMC) fitting method 
with the ability to accurately estimate the probability density function of the redshift for each object.
After the analysis of the redshift 8-12 galaxy sample,
they report that there is an average probability of 21\% that these well defined high redshift galaxies are low redshift interlopers.

Without infrared or deeper HST observations, these EELG contamination can not be completely excluded. 
Therefore, it is quite necessary to properly include the contamination fraction of EELGs for the analysis of high redshift samples.
However, the contamination fraction is still large uncertain due to the lack of observations of EELGs.
The ongoing Hubble Frontier Fields program devote a total 560 orbits to observe
four clusters along with four parallel blank fields. 
This initiative can reach $5\sigma$ magnitude limit of $\sim 28.7$ which is 1.2 deeper than CLASH field.
A systematical selections of EELGs in the Hubble Frontier Fields will help us to constrain the number density of faint EELGs
and resolve the EELG contamination fraction for high redshift galaxies.

\section{Summary}
We have carried out two color-color selections to search for EELGs 
with $EW > 600$ \AA\ in 25 CLASH cluster fields.
We identified two samples consisting of 40 and 12 EELGs, in redshift ranges $1.55\sim1.79$ and $0.93\sim1.14$, respectively.
Thanks to cluster lensing, EELGs are detected down to intrinsic apparent magnitude 28.5,
which is significantly fainter than other samples.
Three candidates are found with extreme $EW>3000$ \AA, which are stronger than other surveys.
We found an abundant population of low-luminosity galaxies whose emission lines are considerably stronger than other samples.
These EELGs are considered as low mass and strong starburst galaxies in the early stage.
Future deep spectral observations are needed for such low luminosity galaxies.

Strong emission lines in these EELGs can boost the broadband photometry by more than one magnitude.
Such extreme emission lines will not only impact the SED fitting, 
but also mimic the dropout feature seen in LBG and contaminate the selections.
We compare the EELGs and the LBG selected from the CLASH data \citep{bradley13},
and find two common objects in our Y sample and the $z \sim 6$ galaxies.
Both the EELGs and the LBGs can explain the spectral type,
but the current photometric data are not deep enough to definitely distinguish between the two scenarios.
We also notice that he possible contamination for future selections of galaxies at $z \sim 10$ cannot be ignored
and the contamination fraction could be significantly higher.
One magnitude deeper observations in the bluer bands ($Y_{105}$ or $YJ_{110}$) or the detection in mid-infrared bands with Spitzer/IRAC
may help us identify the real LBG.

Future deep spectroscopic observations of EELGs like James Webb Space Telescope (JWST) are needed to 
make accurate measurements of emission lines and unveil the nature of these EELGs. 
Furthermore, the ongoing \HST\ observations of frontier fields provides unprecedented deep observations of six clusters
and six parallel fields\footnote{For details, see http://www.stsci.edu/hst/campaigns/frontier-fields/}.
We predict that these deep observations will reveal dozens of EELGs with possible stronger emission lines, 
which will help us to confirm the EW and luminosity trend and estimate the number density.

\textit{Facilities:} HST (ACS,WFC3)
\acknowledgements{
The CLASH program (GO-12065) is based on observations
made with the NASA/ESA {\it Hubble Space Telescope}. The Space Telescope Science
Institute is operated by the Association of Universities for Research in
Astronomy, Inc. under NASA contract NAS 5-26555.
J.X.W. acknowledges support from Chinese Top-notch Young Talents Program and 
the Strategic Priority Research Program “The Emergence of Cosmological Structures" of the Chinese Academy of Sciences (grant No. XDB09000000).
}

\bibliographystyle{apj}

\clearpage

\clearpage

\begin{landscape}
\begin{deluxetable}{cccccccccccccc}
\tabletypesize{\scriptsize}
\setlength{\tabcolsep}{0.05in}
\tablecaption{EELG in the Y Sample}
\tablewidth{0pt}
\tablehead{
 \colhead{EELG}  &\colhead{Cluster} &\colhead{RA     } & \colhead{DEC    } & \colhead{$I_{814}$} & \colhead{$J_{125}$} & \colhead{$H_{160}$}& \colhead{$EW_{YJ110}$ }          & \colhead{$EW_{J125}$ }                & \colhead{$EW_{JH140}$ }                & \colhead{$\beta$} & \colhead{$\mu$} & \colhead{$FWHM$}  & \colhead{log(M)}   \\\\
                 &                  & deg             & deg               & AB             & AB                & AB              &           $\AA$              &   $\AA$                        &   $\AA$                         &                   &                 & arcsec  & $M_{\sun}$ \\
       (1)       &       (2)        &    (3)            & (4)                 & (5)              & (6)                 &(7)               &            (8)                       &   (9)                                 &   (10)                                 &       (11)        &      (12)       & (13)  & (14)\\
}
\startdata
    J1  &            a1423  &   179.33289  &    33.60424  &   25.88 $\pm$ 0.06   &   25.37 $\pm$ 0.05   &   25.99 $\pm$ 0.07   &      640 $\pm$    181   &      774 $\pm$    160   &      842 $\pm$    188   &     1.4 $\pm$  0.1  &   1.6  &  0.37    &  7.8 $\pm $ 0.2 \\ %
    $J2^a$  &             a209  &    22.95528  &   -13.60011  &   27.92 $\pm$ 0.21   &   26.33 $\pm$ 0.10   &   28.21 $\pm$ 0.32   &     3145 $\pm$   1169   &     4259 $\pm$   1354   &     3976 $\pm$   1270   &     2.1 $\pm$  0.4  &   1.9  &  0.29   &  6.7 $\pm $ 6.7 \\
    J3  &             a209  &    22.98728  &   -13.60416  &   26.38 $\pm$ 0.09   &   25.49 $\pm$ 0.05   &   26.67 $\pm$ 0.13   &     1552 $\pm$    383   &     1723 $\pm$    329   &     1312 $\pm$    352   &     2.2 $\pm$  0.1  &   1.2  &  0.25   &  7.2 $\pm $ 0.1 \\
    J4  &             a209  &    22.98267  &   -13.60450  &   25.91 $\pm$ 0.08   &   25.07 $\pm$ 0.05   &   25.73 $\pm$ 0.07   &      983 $\pm$    260   &     1150 $\pm$    219   &     1016 $\pm$    251   &     1.7 $\pm$  0.1  &   1.3  &  0.28   &  7.7 $\pm $ 0.2 \\
    J5  &             a383  &    42.01418  &    -3.54037  &   26.80 $\pm$ 0.11   &   25.92 $\pm$ 0.06   &   26.51 $\pm$ 0.09   &     1569 $\pm$    362   &     1149 $\pm$    269   &     1762 $\pm$    382   &     2.0 $\pm$  0.2  &   1.7  &  0.27   &  6.7 $\pm $ 0.1 \\
    J6  &             a383  &    42.00668  &    -3.54478  &   26.25 $\pm$ 0.07   &   25.68 $\pm$ 0.05   &   26.61 $\pm$ 0.10   &     1215 $\pm$    271   &     1111 $\pm$    214   &      983 $\pm$    264   &     2.4 $\pm$  0.1  &   1.2  &  0.26   &  7.1 $\pm $ 0.1 \\
    J7  &             a383  &    42.00755  &    -3.54529  &   27.10 $\pm$ 0.12   &   26.45 $\pm$ 0.08   &   27.12 $\pm$ 0.13   &     1272 $\pm$    427   &      984 $\pm$    323   &     1255 $\pm$    433   &     2.2 $\pm$  0.2  &   1.2  &  0.30   &  6.9 $\pm $ 0.2 \\
    J8  &        c1226  &   186.75410  &    33.54530  &   27.24 $\pm$ 0.09   &   26.20 $\pm$ 0.10   &   26.95 $\pm$ 0.13   &     1444 $\pm$    421   &     1657 $\pm$    432   &     1353 $\pm$    423   &     2.1 $\pm$  0.2  &   2.7  &  0.41   &  6.7 $\pm $ 0.1 \\ %
    J9  &        c1226  &   186.75931  &    33.53572  &   27.98 $\pm$ 0.15   &   26.90 $\pm$ 0.12   &   27.77 $\pm$ 0.22   &     1615 $\pm$    709   &     1785 $\pm$    634   &     1926 $\pm$    742   &     3.2 $\pm$  0.2  &   1.3  &  0.64   &  6.6 $\pm $ 0.1 \\ %
   J10  &        c1226  &   186.75419  &    33.53426  &   26.68 $\pm$ 0.05   &   25.92 $\pm$ 0.06   &   26.51 $\pm$ 0.11   &                     ... &     1116 $\pm$    211   &     1424 $\pm$    300   &     2.3 $\pm$  0.1  &   1.3  &  0.26   &  7.3 $\pm $ 0.1 \\ %
   J11  &      m0329  &    52.43664  &    -2.20187  &   26.28 $\pm$ 0.09   &   25.64 $\pm$ 0.06   &   26.23 $\pm$ 0.09   &     1218 $\pm$    339   &      946 $\pm$    245   &     1041 $\pm$    296   &     2.5 $\pm$  0.1  &   2.8  &  0.77   &  6.9 $\pm $ 0.2 \\
   J12  &      m0329  &    52.42710  &    -2.20908  &   26.39 $\pm$ 0.09   &   25.70 $\pm$ 0.06   &   26.75 $\pm$ 0.12   &     1720 $\pm$    393   &     1401 $\pm$    297   &     1522 $\pm$    352   &     2.0 $\pm$  0.1  &   2.9  &  0.28   &  6.7 $\pm $ 0.1 \\
   J13  &      m0416  &    64.04457  &   -24.07828  &   24.65 $\pm$ 0.04   &   23.83 $\pm$ 0.02   &   24.53 $\pm$ 0.04   &      801 $\pm$    127   &     1198 $\pm$    110   &     1052 $\pm$    126   &     1.7 $\pm$  0.1  &   2.0  &  0.25   &  7.7 $\pm $ 0.1 \\
   J14  &      m0416  &    64.04506  &   -24.08220  &   26.51 $\pm$ 0.11   &   25.58 $\pm$ 0.06   &   26.35 $\pm$ 0.11   &     1569 $\pm$    417   &     1398 $\pm$    327   &     1252 $\pm$    368   &     1.7 $\pm$  0.3  &   1.4  &  0.27   &  7.1 $\pm $ 0.1 \\
   J15  &      m0429  &    67.39312  &    -2.88310  &   24.52 $\pm$ 0.03   &   23.68 $\pm$ 0.02   &   24.40 $\pm$ 0.03   &     1560 $\pm$    122   &     1268 $\pm$     98   &     1495 $\pm$    119   &     2.0 $\pm$  0.1  &   2.4  &  0.72   &  7.6 $\pm $ 0.1 \\
   J16  &       m0647  &   101.89789  &    70.24444  &   26.75 $\pm$ 0.06   &   25.97 $\pm$ 0.06   &   26.79 $\pm$ 0.13   &                     ... &     1413 $\pm$    259   &     1428 $\pm$    342   &     1.9 $\pm$  0.2  &   1.6  &  0.40   &  6.7 $\pm $ 0.1 \\
   J17  &       m0717  &   109.36207  &    37.74810  &   26.41 $\pm$ 0.06   &   25.70 $\pm$ 0.09   &   26.76 $\pm$ 0.12   &     1652 $\pm$    350   &     1356 $\pm$    312   &     1309 $\pm$    279   &     2.4 $\pm$  0.1  &   2.1  &  0.30   &  6.8 $\pm $ 0.1 \\
   J18  &       m0744  &   116.22197  &    39.44121  &   27.13 $\pm$ 0.12   &   25.95 $\pm$ 0.08   &   26.79 $\pm$ 0.14   &     1630 $\pm$    554   &     2056 $\pm$    501   &     1203 $\pm$    444   &     2.8 $\pm$  0.2  &   1.2  &  0.88   &  6.6 $\pm $ 0.1 \\
   J19  &       m1115  &   168.97679  &     1.51242  &   27.48 $\pm$ 0.18   &   26.43 $\pm$ 0.09   &   27.65 $\pm$ 0.21   &     1873 $\pm$    767   &     2108 $\pm$    668   &     2394 $\pm$    814   &     2.1 $\pm$  0.3  &   1.4  &  0.24   &  6.5 $\pm $ 0.1 \\
   J20  &       m1115  &   168.97415  &     1.50723  &   26.56 $\pm$ 0.10   &   25.70 $\pm$ 0.06   &   26.96 $\pm$ 0.15   &     1849 $\pm$    463   &     1856 $\pm$    388   &     2022 $\pm$    462   &     2.4 $\pm$  0.2  &   1.7  &  0.24   &  6.7 $\pm $ 0.1 \\
   J21  &       m1115  &   168.95223  &     1.50104  &   25.38 $\pm$ 0.05   &   24.70 $\pm$ 0.04   &   25.31 $\pm$ 0.05   &     1072 $\pm$    166   &      972 $\pm$    132   &      944 $\pm$    158   &     2.3 $\pm$  0.1  &   2.0  &  0.39   &  7.8 $\pm $ 0.1 \\
   J22  &       m1149  &   177.39721  &    22.40619  &   25.59 $\pm$ 0.06   &   24.51 $\pm$ 0.03   &   25.80 $\pm$ 0.08   &     2516 $\pm$    293   &     2214 $\pm$    245   &     2174 $\pm$    276   &     2.1 $\pm$  0.1  &   8.4  &  0.31   &  6.2 $\pm $ 0.1 \\
   J23  &      m1206  &   181.55219  &    -8.78764  &   26.62 $\pm$ 0.12   &   25.54 $\pm$ 0.05   &   26.45 $\pm$ 0.11   &     1390 $\pm$    381   &     1823 $\pm$    358   &     1264 $\pm$    369   &     2.2 $\pm$  0.2  &   1.8  &  0.40   &  6.8 $\pm $ 0.1 \\
   J24  &      m1206  &   181.56703  &    -8.81022  &   26.47 $\pm$ 0.08   &   25.88 $\pm$ 0.06   &   27.34 $\pm$ 0.26   &                     ... &     1299 $\pm$    302   &     1634 $\pm$    423   &     2.2 $\pm$  0.1  &   2.7  &  0.31   &  6.7 $\pm $ 0.1 \\
   J25  &    m1311  &   197.74210  &    -3.16371  &   25.55 $\pm$ 0.05   &   24.71 $\pm$ 0.03   &   25.23 $\pm$ 0.06   &                     ... &     1161 $\pm$    150   &     1233 $\pm$    195   &     2.0 $\pm$  0.1  &   1.2  &  0.25   &  7.5 $\pm $ 0.1 \\  %
   J26  &    m1311  &   197.77449  &    -3.16473  &   27.19 $\pm$ 0.15   &   26.22 $\pm$ 0.09   &   26.91 $\pm$ 0.19   &                     ... &     1461 $\pm$    493   &     1786 $\pm$    656   &     2.3 $\pm$  0.2  &   1.2  &  0.28   &  6.9 $\pm $ 0.2 \\  %
   J27  &    m1311  &   197.77126  &    -3.16271  &   26.02 $\pm$ 0.08   &   25.25 $\pm$ 0.06   &   25.91 $\pm$ 0.12   &                     ... &     1133 $\pm$    250   &     1185 $\pm$    320   &     2.6 $\pm$  0.1  &   1.3  &  0.45   &  7.3 $\pm $ 0.2 \\  %
   J28  &       m1423  &   215.95701  &    24.09266  &   26.52 $\pm$ 0.06   &   25.76 $\pm$ 0.06   &   27.00 $\pm$ 0.15   &      846 $\pm$    299   &     1442 $\pm$    275   &     1257 $\pm$    308   &     2.0 $\pm$  0.1  &   2.2  &  0.57   &  6.8 $\pm $ 0.2 \\
   $J29^a$  &        m1720  &   260.05902  &    35.62797  &   27.08 $\pm$ 0.11   &   25.65 $\pm$ 0.04   &   27.06 $\pm$ 0.18   &                     ... &     3145 $\pm$    555   &     3648 $\pm$    718   &     2.0 $\pm$  0.2  &   1.1  &  0.25   &  6.6 $\pm $ 0.1 \\
   J30  &       m1720  &   260.07253  &    35.62020  &   27.35 $\pm$ 0.14   &   26.18 $\pm$ 0.07   &   27.32 $\pm$ 0.17   &     1209 $\pm$    501   &     2207 $\pm$    536   &     2191 $\pm$    528   &     2.2 $\pm$  0.2  &   2.1  &  0.23   &  6.9 $\pm $ 0.1 \\
   J31  &      m1931  &   292.93714  &   -26.58000  &   27.30 $\pm$ 0.17   &   26.50 $\pm$ 0.10   &   27.50 $\pm$ 0.19   &     1974 $\pm$    710   &     1524 $\pm$    544   &     1266 $\pm$    589   &     2.3 $\pm$  0.3  &   1.2  &  0.31   &  7.0 $\pm $ 0.2 \\
   J32  &      m2129  &   322.36038  &    -7.67390  &   25.98 $\pm$ 0.06   &   25.16 $\pm$ 0.05   &   25.69 $\pm$ 0.06   &     1454 $\pm$    248   &     1162 $\pm$    197   &     1274 $\pm$    230   &     2.1 $\pm$  0.1  &   1.2  &  0.29   &  7.5 $\pm $ 0.1 \\ %
   J33  &      m2129  &   322.36494  &    -7.70077  &   27.10 $\pm$ 0.11   &   26.33 $\pm$ 0.09   &   27.11 $\pm$ 0.14   &     1121 $\pm$    450   &     1271 $\pm$    381   &     1239 $\pm$    433   &     2.0 $\pm$  0.2  &   1.4  &  0.46   &  7.1 $\pm $ 0.2 \\ %
   J34  &      m2129  &   322.37282  &    -7.70210  &   26.74 $\pm$ 0.08   &   26.07 $\pm$ 0.12   &   27.42 $\pm$ 0.23   &     1059 $\pm$    376   &     1386 $\pm$    433   &     1611 $\pm$    500   &     2.4 $\pm$  0.1  &   1.3  &  0.30   &  7.1 $\pm $ 0.2 \\ %
   J35  &          m2137  &   325.06561  &   -23.64302  &   25.73 $\pm$ 0.06   &   25.06 $\pm$ 0.04   &   25.87 $\pm$ 0.06   &     1339 $\pm$    209   &     1166 $\pm$    172   &     1314 $\pm$    209   &     1.4 $\pm$  0.1  &   1.6  &  0.33   &  7.4 $\pm $ 0.1 \\
   J36  &          m2137  &   325.04950  &   -23.67534  &   25.81 $\pm$ 0.06   &   25.02 $\pm$ 0.04   &   25.74 $\pm$ 0.06   &      862 $\pm$    190   &     1180 $\pm$    174   &     1242 $\pm$    207   &     1.2 $\pm$  0.1  &   1.4  &  0.27   &  7.7 $\pm $ 0.1 \\
   J37  &          m2137  &   325.06148  &   -23.67546  &   25.64 $\pm$ 0.06   &   24.95 $\pm$ 0.04   &   25.53 $\pm$ 0.05   &      949 $\pm$    174   &      942 $\pm$    146   &     1020 $\pm$    177   &     2.0 $\pm$  0.1  &   1.5  &  0.38   &  7.9 $\pm $ 0.1 \\
   J38  &       r1347  &   206.90479  &   -11.75055  &   23.73 $\pm$ 0.04   &   22.97 $\pm$ 0.01   &   23.60 $\pm$ 0.01   &      800 $\pm$     49   &      826 $\pm$     44   &      912 $\pm$     49   &     1.8 $\pm$  0.1  &   1.4  &  0.56   &  8.5 $\pm $ 0.1 \\
   J39  &        r1532  &   233.22674  &    30.33152  &   26.79 $\pm$ 0.14   &   25.87 $\pm$ 0.08   &   26.55 $\pm$ 0.12   &     1127 $\pm$    440   &     1265 $\pm$    372   &     1243 $\pm$    436   &     2.1 $\pm$  0.2  &   1.4  &  0.45   &  7.2 $\pm $ 0.2 \\  %
   J40  &       r2248  &   342.17207  &   -44.51618  &   25.80 $\pm$ 0.04   &   24.93 $\pm$ 0.03   &   25.90 $\pm$ 0.06   &     1615 $\pm$    192   &     1467 $\pm$    164   &     1861 $\pm$    207   &     1.8 $\pm$  0.1  &   1.6  &  0.59   &  7.2 $\pm $ 0.1 \\

\tablenotetext{a}{EELGs which have error weighted average of EWs higher than 3000 \AA. }
\tablecomments{
(1) identification for each EELG.
(2) the short names of cluster fields for each EELG: 
a1423, Abell 1423; a209, Abell 209; a383, Abell 383; a611, Abell 611; a2261, Abell 2261; c1226, CLJ1226.9+3332; m0329, MACS0329.7--0211; m0416, MACS0416.1--2403; m0429, MACS0429.6--0253; m0647, MACS0647.8+7015; m0717, MACS0717.5+3745; m0744, MACS0744.9+3927; m1115, MACS1115.9+0129; m1149, MACS1149.6+2223; m1206, MACS1206.2--0847; m1311, MACS1311.0--0310; m1423, MACS1423.8+2404; m1720, MACS1720.3+3536; m1931, MACS1931.8--2635; m2129, MACS2129.4--0741; m2137, MS2137--2353; r1347, RXJ1347.5--1145; r1532, RXJ1532.9+3021; r2129, RXJ2129.7+0005; r2248, RXJ2248.7--4431.
(3)-(4) coordinates in J2000.
(5)-(7) apparent magnitudes for $I_{814}$, $J_{125}$, $H_{160}$.
(8)-(10) EWs derived from the flux excesses in $YJ_{110}$, $J_{125}$ and $JH_{140}$.
(11) continuous slopes inferred from linear fits to the ACS/WFC bands.
(12) magnifications estimated from the lensing models.
(13) the apparent sizes in $J_{125}$ images derived from {\tt SExtractor} in unit of arcsec. 
(14) stellar masses inferred from Starburst99 \citet{leitherer99}.
}
\label{table1}
\end{deluxetable}
\clearpage
\end{landscape}

\begin{landscape}
\begin{deluxetable}{cccccccccccccc}
\tabletypesize{\scriptsize}
\setlength{\tabcolsep}{0.05in}
\tablecaption{EELG in the Y Sample}
\tablewidth{0pt}
\tablehead{
 \colhead{EELG} &\colhead{Cluster} &\colhead{RA}    & \colhead{DEC}       & \colhead{$I_{814}$} & \colhead{$Y_{105}$}    & \colhead{$H_{160}$} & \colhead{$EW_{Y105}$ } & \colhead{$EW_{YJ110}$ } & \colhead{$EW_{J125}$ }& \colhead{$\beta$}  & \colhead{$\mu$} & \colhead{FWHM}   & \colhead{log(M)} \\
                 &                  & deg             & deg               & AB             & AB                & AB              &               $\AA$                        &   $\AA$                      &          $\AA$                      &               &                 & arcsec    & $M_{\sun}$  \\
       (1)       &       (2)        &    (3)            & (4)                 & (5)              & (6)                 &(7)               &            (8)                &   (9)                          &   (10)                       &       (11)                             &      (12)          & (13)       & (14)     \\
}
 \startdata
  Y1  &           a383  &    42.00382  &    -3.53457  &   26.46 $\pm$ 0.08   &   25.85 $\pm$ 0.09   &   26.51 $\pm$ 0.09   &     1108 $\pm$    332   &      985 $\pm$    346   &      843 $\pm$    269   &                         2.1 $\pm$  0.1  &   1.4  &  0.37   &  7.4 $\pm $ 0.3 \\
    Y2  &             a383  &    42.01025  &    -3.53787  &   25.68 $\pm$ 0.05   &   24.76 $\pm$ 0.04   &   25.33 $\pm$ 0.04   &     1387 $\pm$    182   &     1220 $\pm$    194   &     1310 $\pm$    157   &        2.2 $\pm$  0.1  &   2.0  &  0.28   &  6.7 $\pm $ 0.1 \\
   Y3  &          a383  &    42.00173  &    -3.54142  &   25.08 $\pm$ 0.03   &   24.59 $\pm$ 0.03   &   25.39 $\pm$ 0.04   &     1085 $\pm$    131   &     1501 $\pm$    156   &      415 $\pm$     97   &         1.7 $\pm$  0.1  &   1.2  &  0.25   &  7.3 $\pm $ 0.1 \\
   Y4  &         c1226  &   186.75661  &    33.54715  &   26.55 $\pm$ 0.06   &   25.62 $\pm$ 0.06   &   26.17 $\pm$ 0.08   &     1644 $\pm$    290   &     1797 $\pm$    380   &     1640 $\pm$    320   &         1.7 $\pm$  0.2  &   1.2  &  0.26   &  6.4 $\pm $ 0.1 \\  %
   Y5  &         m0329  &    52.44139  &    -2.20040  &   27.10 $\pm$ 0.15   &   26.15 $\pm$ 0.08   &   27.16 $\pm$ 0.16   &     2072 $\pm$    606   &     1022 $\pm$    674   &     1340 $\pm$    545   &                        2.0 $\pm$  0.3  &   1.5  &  0.38   &  7.0 $\pm $ 0.2 \\
    Y6      &      m0329  &    52.41057  &    -2.20123  &   26.62 $\pm$ 0.10   &   26.00 $\pm$ 0.07   &   26.70 $\pm$ 0.11   &     1173 $\pm$    343   &     1217 $\pm$    454   &      695 $\pm$    313   &          2.4 $\pm$  0.2  &   1.6  &  0.43   &  6.4 $\pm $ 0.3 \\
    Y7      &       m0717  &   109.40992  &    37.76028  &   26.10 $\pm$ 0.04   &   25.45 $\pm$ 0.06   &   26.66 $\pm$ 0.15   &     1399 $\pm$    264   &     1897 $\pm$    349   &                     ... &         1.7 $\pm$  0.1  &   1.8  &  0.25   &  6.5 $\pm $ 0.1 \\
    $Y8^a$  &      m1115  &   168.97197  &     1.51001  &   27.89 $\pm$ 0.26   &   26.96 $\pm$ 0.13   &   27.80 $\pm$ 0.25   &     1754 $\pm$    910   &     1626 $\pm$   1226   &      230 $\pm$    746   &            ...        &   1.5  &  0.42   &  6.3 $\pm $ 0.7 \\
    $Y9^a$  &       m1720  &   260.08484  &    35.60698  &   28.24 $\pm$ 0.28   &   26.81 $\pm$ 0.11   &   27.96 $\pm$ 0.26   &     3307 $\pm$   1293   &     3165 $\pm$   1459   &     2059 $\pm$   1099   &            ...        &   1.2  &  0.45   &  6.1 $\pm $ 0.1 \\
   Y10  &       m2129  &   322.35663  &    -7.68367  &   26.85 $\pm$ 0.10   &   25.93 $\pm$ 0.07   &   26.59 $\pm$ 0.09   &     1562 $\pm$    373   &     1598 $\pm$    472   &     1778 $\pm$    404   &                      2.9 $\pm$  0.2  &   1.5  &  0.28   &  7.0 $\pm $ 0.1 \\   %
   Y11      &      m2129  &   322.33898  &    -7.69112  &   23.67 $\pm$ 0.01   &   23.13 $\pm$ 0.02   &   24.19 $\pm$ 0.04   &     1101 $\pm$     63   &                     ... &      375 $\pm$     51   &        1.8 $\pm$  0.1  &   1.5  &  0.54   &  8.0 $\pm $ 0.1 \\  %
   Y12      &       r1347  &   206.87554  &   -11.74331  &   26.69 $\pm$ 0.07   &   26.06 $\pm$ 0.08   &   27.01 $\pm$ 0.10   &     1434 $\pm$    343   &     1647 $\pm$    349   &     1211 $\pm$    356   &       2.2 $\pm$  0.1  &   3.4  &  0.33   &  6.2 $\pm $ 0.1 \\

\enddata

\tablenotetext{a}{Candidates which can be explained with both redshift $5-6$ galaxies and strong emission lines. The Y8 and Y9 are selected as MACS1115-0352 and MACS1720-1114 in the $z \sim 6$ catalog of \citet{bradley13}.}
\tablecomments{As Table \ref{table1}. But the (5)-(7) are apparent magnitudes for $I_{814}$, $Y_{105}$ , $H_{160}$ and (8)-(10) are EWs derived from the flux excesses in $YJ_{110}$, $J_{125}$ and $JH_{140}$. }
\label{table2}
\end{deluxetable}
\clearpage

\end{landscape}

\end{document}